%% file: main.tex
\documentclass[sigconf]{acmart}
\usepackage{booktabs,multirow}
\usepackage{bbold}
\usepackage[linesnumbered,algoruled,boxed,lined]{algorithm2e}
\usepackage{array}
\usepackage{efbox}
\usepackage{enumitem}
\usepackage{setspace}

\newcommand{\ie}{{\it i.e.}}
\newcommand{\eg}{{\it e.g.}}

\newcommand{\topN}{top-\emph{N}}
\newcommand{\argmin}{\operatornamewithlimits{argmin}}

\copyrightyear{2021}
\acmYear{2021}
\setcopyright{acmcopyright}
\acmConference[WSDM '21]{Proceedings of the Fourteenth ACM International Conference on Web Search and Data Mining}{March 8--12, 2021}{Virtual Event, Israel}
\acmBooktitle{Proceedings of the Fourteenth ACM International Conference on Web Search and Data Mining (WSDM '21), March 8--12, 2021, Virtual Event, Israel}
\acmPrice{15.00}
\acmDOI{10.1145/3437963.3441808}
\acmISBN{978-1-4503-8297-7/21/03}

\begin{document}
\fancyhead{}

\title{Local Collaborative Autoencoders}

\settopmatter{authorsperrow=4}
\author{Minjin Choi}
\affiliation{
  \institution{Sungkyunkwan University}
  \institution{Republic of Korea}}
\email{zxcvxd@skku.edu}

\author{Yoonki Jeong}
\affiliation{
  \institution{Sungkyunkwan University}
  \institution{Republic of Korea}}  
\email{yoongi0428@skku.edu}

\author{Joonseok Lee}
\affiliation{
  \institution{Google Research}
  \institution{United States, CA}}  
\email{joonseok@google.com}

\author{Jongwuk Lee}\authornote{Corresponding author}
\affiliation{
  \institution{Sungkyunkwan University}
  \institution{Republic of Korea}}  
\email{jongwuklee@skku.edu}

\input{sec-abstract}
\maketitle

\input{sec-introduction}
\input{sec-preliminaries}
\input{sec-model}

\input{sec-experiments}

\input{sec-results}

\input{sec-relatedwork}
\input{sec-conclusion}

\bibliographystyle{ACM-Reference-Format}
\bibliography{references}

\input{sec-appendix}

\end{document}

%% file: sec-abstract.tex
\begin{abstract}
Top-$N$ recommendation is a challenging problem because complex and sparse user-item interactions should be adequately addressed to achieve high-quality recommendation results. The local latent factor approach has been successfully used with multiple local models to capture diverse user preferences with different sub-communities. However, previous studies have not fully explored the potential of local models, and failed to identify many small and coherent sub-communities. In this paper, we present \emph{Local Collaborative Autoencoders (LOCA)}, a generalized local latent factor framework. Specifically, LOCA adopts different neighborhood ranges at the training and inference stages. Besides, LOCA uses a novel sub-community discovery method, maximizing the coverage of a union of local models and employing a large number of diverse local models. By adopting autoencoders as the base model, LOCA captures latent non-linear patterns representing meaningful user-item interactions within sub-communities. Our experimental results demonstrate that LOCA is scalable and outperforms state-of-the-art models on several public benchmarks, by 2.99--4.70\% in Recall and 1.02--7.95\% in NDCG, respectively.

\end{abstract}

\ccsdesc[500]{Information systems~Collaborative filtering}
\ccsdesc[300]{Information systems~Expert systems}

\keywords{Collaborative filtering; local latent factor model; autoencoders}

%% file: sec-introduction.tex
\section{Introduction}\label{sec:introduction}

\input{Fig1_example.tex}

Neural recommender models~\cite{ZhangYS17aa,SedhainMSX15,WuDZE16,WangWY15, Cheng0HSCAACCIA16,CovingtonAS16,HeDWTTC18,KimPOLY16,HidasiKBT15,WuABSJ17,LeeAVN18,Lee20gcml,ChenR18,LobelLGC19,HeHSLJC18,Wang0WFC19,ShenbinATMN20} have been actively studied for representing complex and non-linear factors across user-item interactions. Although deep neural networks (DNNs) are generally capable of identifying the complex patterns of user-item interactions, recent studies~\cite{DacremaCJ19,LudewigMLJ19} reported that the performance gain in the recommendation problem is less path-breaking than in other domains such as computer vision, speech recognition, or natural language processing. There may be multiple reasons, such as extreme data sparsity, data ambiguity for missing feedback, and noisy user feedback. However, it is still an open question as to why DNNs have achieved less satisfactory improvements in the recommendation problem.

In this paper, we hypothesize the reason for the above as follows: DNNs adopt a deep network architecture and non-linear activation layers, and hence, are generally more potent than traditional models; however, neural recommender models usually employ shallow networks owing to the sparsity of training data. Such a shallow architecture with traditional training schemes alone is limited in identifying diverse local patterns of user-item interactions and are mostly biased to the global pattern.

To address this problem, we revisit the underlying assumption that the rating matrix is \emph{globally} low-rank. \citet{LeeKLS13, LeeKLSB16} proposed a \emph{local low-rank assumption} (Figure~\ref{fig:example}), significantly relaxing the global counterpart, according to which: ``A rating matrix is not necessarily decomposed by global low-rank matrices, and it is a union of multiple local low-rank matrices composed of a subset of users/items that share local interests.'' For example, in the movie domain, a sub-group of users preferring romantic comedy movies is in low-rank; another sub-group of users who prefer science fiction films in the 2010s is also in low-rank; however, a union of these may not be in low-rank. We can thus presume that the dominant factors determining the preference of each group are quite different in these two sub-groups, \emph{e.g.}, the quality of computer graphics, which may matter only in the latter. Based on this intuition, the local low-rank assumption has been successfully applied to the rating prediction~\cite{LeeKLS13, LeeKLSB16} and the ranking problem~\cite{LeeBKLS14}, with WMF~\cite{WangPJSW16}, SLIM~\cite{Christakopoulou16}, and SVD~\cite{Christakopoulou18} as its base models.

However, the existing local models did not fully explore the potential of the local low-rank assumption for the following three aspects. First, the existing studies~\cite{LeeKLS13,LeeBKLS14,WangPJSW16,Christakopoulou16,Christakopoulou18} did not fully discover diverse locality. Adopting a relatively broad coverage of users in the local model~\cite{LeeKLS13,LeeBKLS14,WangPJSW16}, the size of the local models is close to that of the global model, thereby violating the intuition of local models. Although~\citet{Christakopoulou16,Christakopoulou18} employed the local models with small sub-communities, their performance was often worse than that of the global model owing to the lack of training data. Second, existing local models did not attempt to train and combine a large number of local models (\ie, at most 100 models) to capture smaller and more coherent sub-communities. Thus, the previous studies can be regarded as developing a \emph{small} number of relatively \emph{similar} models; their performance improvement is most likely due to the ensemble effect, similar to bagging~\cite{LeeKLSB16}. Lastly, the base models for learning the local models were often limited to linear latent factor models, which made it difficult to identify meaningful non-linear patterns in local models.

In this paper, we propose a novel local recommender framework, called \emph{Local Collaborative Autoencoders} (\emph{LOCA}). The key novelty of LOCA is as follows: (i) It provides a generalized architecture for learning a variety of local models by identifying various sub-communities for training and inference; most of the previous work on local models can be considered to be the special cases of LOCA. (ii) LOCA can handle a large number of small and coherent sub-communities for the local model; thus, its performance is improved with more local models beyond the level explored in the previous studies. (iii) Although the architecture of LOCA is model-agnostic for base models, we utilize the autoencoder-based model as its base model. Autoencoder-based models have non-linear activation layers to represent the meaningful non-linear patterns in the local model. To the best of our knowledge, it is the first attempt that adopts the local low-rank assumption using neural recommender models. In experimental results, LOCA achieves ground-breaking performance, significantly outperforming state-of-the-art global and local models on five public benchmark datasets.

%% file: Fig1_example.tex
\begin{figure}
  \includegraphics[width=0.32\textwidth]{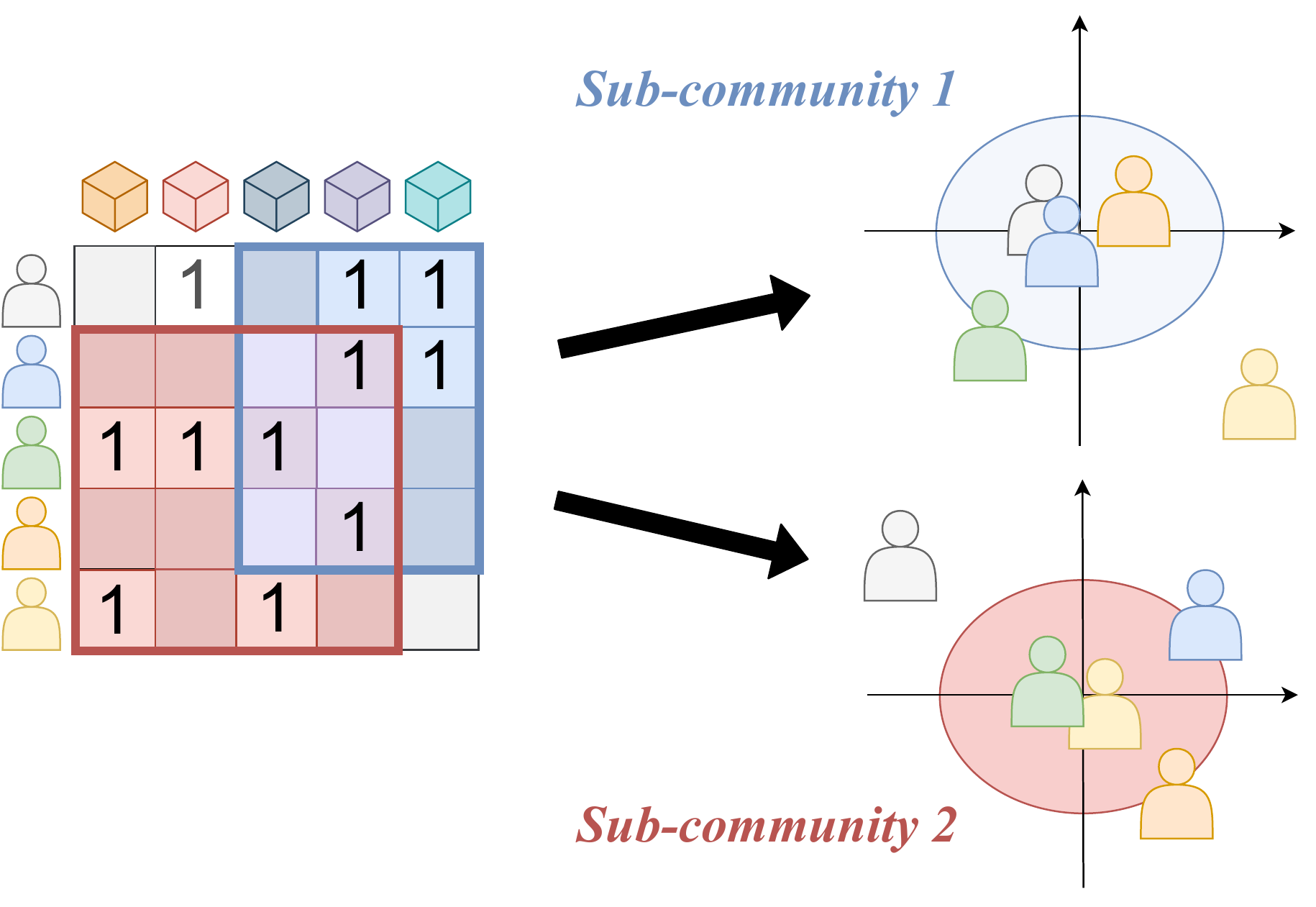}
  \caption{Illustration of the local low-rank assumption. Two sub-matrices are locally low-rank, in which one can discover different sub-communities.}
  \label{fig:example}
  \vskip -0.18in
\end{figure}

%% file: sec-preliminaries.tex
\section{Preliminaries}\label{sec:preliminaries}

Given a set $\mathcal{U}$ of $m$ users and a set $\mathcal{I}$ of $n$ items, we have a binary rating matrix $\mathbf{R} \in \{0, 1\}^{m \times n}$. An entry $r_{ui} \in \mathbf{R}$ represents an implicit feedback by user $u \in \mathcal{U}$ on item $i \in \mathcal{I}$. If $r_{ui} = 1$, it indicates a positive feedback; otherwise, it indicates a \emph{missing} (or \emph{unobserved}) feedback. Given a user $u$, $\mathcal{I}_u^+ = \{i \in \mathcal{I} | r_{ui} = 1\}$ and $\mathcal{I}_u^- = \mathcal{I} - \mathcal{I}_u^+$ are a set of items with positive and missing feedback, respectively.

Our goal is to retrieve a ranked list of the \topN\ items that the user $u$ prefers the most. Given a rating matrix $\mathbf{R}$, we first learn a recommendation model $M(\mathbf{R}; \theta): \{0, 1\}^{m \times n} \rightarrow \mathbb{R}^{m \times n}$ with parameter $\theta$ to infer a preference score $\hat{r}_{ui}$ for user $u$ on item $i \in \mathcal{I}_u^-$. We use the same notation for a single user, \ie, $M(\mathbf{r}; \theta):$ $\{0, 1\}^{n}$ $\rightarrow$ $\mathbb{R}^{n}$, where $\mathbf{r}$ is a row of $\mathbf{R}$.

\subsection{Local Latent Factor Models}\label{sec:review_local}

When user preferences are locally coherent but globally diverse, the global model is insufficient to represent different local factors of the users. To address this problem, \citet{LeeKLS13} proposed the \emph{local low-rank assumption}, where the global matrix may be in high rank but is a union of multiple low-rank sub-matrices composed of rows and columns with similar taste. It significantly relaxes the previous assumption that the matrix is globally at low-rank. Subsequent to the pioneering work, namely local low-rank matrix approximation (LLORMA) for rating prediction~\cite{LeeKLS13,LeeKLSB16}, the local low-rank assumption was applied to follow-up works, either by allowing overlaps between local sub-matrices~\cite{LeeBKLS14,WangPJSW16} or by splitting the entire matrix into disjoint multiple sub-matrices~\cite{Christakopoulou16, Christakopoulou18}.



A local model $M^{\text{local}}(\mathbf{R}; \theta^{(j)})$ is trained with its corresponding weight. Formally, each local model has the corresponding weight $\mathcal{T} = \{\mathbf{T}^{(1)}, \dots, \mathbf{T}^{(q)}\}$, where $\mathbf{T}^{(j)} \in \mathbb{R}^{m \times n}$ (for $j = 1, ..., q$) represents the importance (or weights) of each (user, item) pair to the rating matrix $\mathbf{R}$. After all the local models are trained, $\mathbf{R}$ is approximated by aggregating the multiple local models:
\begin{equation}
  \mathbf{\hat{R}} = \sum_{j=1}^{q}{\mathbf{T}^{(j)} \odot M^{\text{local}}(\mathbf{R}; \theta^{(j)}) \oslash \mathbf{T}},
\end{equation}
where, $\mathbf{T} = \sum_{j=1}^{q}{\mathbf{T}^{(j)}}$, and $\odot$ and $\oslash$ are the element-wise product and division, respectively.

\citet{Christakopoulou16, Christakopoulou18} proposed a method which integrates the global model with multiple local models on top of SLIM~\cite{NingK11} and SVD~\cite{CremonesiKT10}. Let $M^{\text{global}}(\mathbf{R}; \theta^{(g)})$ denote the global model.
\begin{equation}
  \mathbf{\hat{R}} = \alpha M^{\text{global}}(\mathbf{R}; \theta^{(g)}) + (1 - \alpha)\sum_{j=1}^{q}{\mathbf{T}^{(j)} \odot M^{\text{local}}(\mathbf{R}; \theta^{(j)}) \oslash \mathbf{T}},
\end{equation}
where $\alpha$ is a hyper-parameter to control the importance of the global model.

\subsection{Autoencoder-based Models}

Recently, autoencoders (AEs) have been adopted as a neural item-to-item recommender models~\cite{SedhainMSX15,Strub15,WuDZE16,LiangKHJ18,Steck19}. The goal of an AE is to minimize the reconstruction error for a set of user rating vectors:
\begin{equation}
  \argmin_{\theta}\sum_{u=1}^m {\mathcal{L} \left(\mathbf{r}_u, M(\mathbf{r}_u; \theta) \right)} + \lambda \Omega(\theta),
\end{equation}
where $\mathbf{r}_u = (r_{u1}, \dots, r_{un})$ is the $u$-th user vector of $\mathbf{R}$, $\lambda$ is the regularization coefficient, and $\Omega(\cdot)$ is a regularization function. Conventional AE-based models~\cite{SedhainMSX15,WuDZE16} have been widely used the cross-entropy loss function  $\mathcal{L}_{CE}$ as the surrogate function for the \topN\ recommendation problem:
\begin{equation}
\mathcal{L}_{CE}(\mathbf{r}_u, \mathbf{\hat{r}}_u) = - \sum_{i=1}^{n}{r_{ui}\log \hat{r}_{ui} + (1 - r_{ui})\log(1 - \hat{r}_{ui})},
\end{equation}
where $\mathbf{r}_u, \hat{\mathbf{r}}_u \in \mathbb{R}^n$ are the input rating vector and the predicted rating vector of user $u$, respectively.

As the variants of AE, denoising autoencoders (DAEs)~\cite{VincentLBM08} and variational autoencoders (VAEs)~\cite{KingmaW13} were used for \topN\ recommendation. CDAE~\cite{WuDZE16} utilizes a DAE by corrupting the input vector randomly. MultVAE~\cite{LiangKHJ18} and RecVAE~\cite{ShenbinATMN20} applied the VAE for \topN\ recommendation. As the simplified version of SLIM~\cite{NingK11}, EASE$^{R}$~\cite{Steck19} showed state-of-the-art results. However, the existing AE-based models focused only on developing a single global model without taking the locality in the data into account.

%% file: sec-model.tex
\section{Proposed Model}
\label{sec:proposed}


\subsection{Motivation}\label{sec:motivation}

Our primary goal is to build multiple local models that represent small and coherent sub-communities. The existing studies~\cite{LeeKLS13,LeeBKLS14, WangPJSW16,Christakopoulou16,Christakopoulou18} developed local models for this purpose; however, they have limited potential. LLORMA and its variants~\cite{LeeKLS13, LeeBKLS14, WangPJSW16} showed superior performance over a global model, but their local models covered a broad range of an entire matrix, where the local models with the wide coverage were comparable to the global model. As discussed in~\cite{LeeKLSB16}, the improvement of LLORMA is statistically significant over an ensemble method; however, it is also true that the improvement mostly comes from the ensemble effect, not just from the characteristics of local models. Moreover, training local models with small sub-matrices in~\cite{Christakopoulou16, Christakopoulou18} suffers from the lack of training data, resulting in low accuracy; The global model largely compensates for the lack of training data issues in local models, which does not adhere to the effect of local models.

\input{Fig2_similarity.tex}

To address this problem, we hypothesize that the optimal size of the neighborhood can be different in the training and inference stages. This is because training, by nature, tends to benefit from more extensive data, whereas relevance may play a more critical role in inference. In accordance with the conventional nearest-neighbor approaches~\cite{GoldbergNOT92, HerlockerKBR99, SarwarKKR01}, we utilize more neighbors for training and focus on a smaller neighborhood with strongly connected target users for inference. This strategy can be regarded as data augmentation to improve the training of local models, representing small and coherent sub-communities. However, unlike existing data augmentation that synthetically generates \emph{virtual} users, we choose \emph{real} users from the neighborhood in the entire matrix.

We explore it with an empirical study shown in Figure~\ref{fig:all_data_sim}. Suppose that we train a local model centered on a particular anchor user, considering the top-1\% of its closest neighbors. When we infer the basis of the same neighborhood as in previous works~\cite{LeeKLS13,LeeBKLS14, WangPJSW16}, the accuracy of the local model (the diamond point at the bottom left) is much worse than that of the global model, implying that the training data is insufficient to learn the local model. Then, maintaining the same neighborhood in the inference step, we expand the neighborhood in the training step to 5\%, 10\%, 20\%, and so on. We observe that the accuracy of the local model improves as the size of the training data increases up to 20\%, even if the augmented users are less similar to the anchor user. From this pilot study, it is clear that the number of neighbors in the training step tends to appreciate at the cost of reduced similarity, whereas the inference step maintains a small number of neighbors for the small and coherent sub-community.

This empirical study motivates us to adopt different schemes for training and inferring local models; That is, when training local models, it is necessary to utilize a broader range of the sub-matrix. By contrast, each local model covers only a small user subset that represents coherent sub-communities during inference, effectively reflecting the intuition of local models.

\subsection{The LOCA Framework}

Figure~\ref{fig:Overview} illustrates the proposed framework, namely \emph{Local Collaborative Autoencoders (LOCA)}. It should be noted that we have described our framework by focusing on users only, although the same logic can be applied to the item-side as in~\cite{LeeKLS13}. Because we address the top-$N$ item recommendation problem, it is of less practical interest to choose the top-$N$ users per item.

Similar to LLORMA~\cite{LeeKLS13,LeeKLSB16}, our model is based on a divide-and-conquer approach with three steps: (i) discovering a set of local communities (\emph{divide}), (ii) training a local model per each sub-community (\emph{conquer}), and (iii) inferring user preferences by combining the global model and multiple local models (\emph{aggregation}).

\input{Fig3_Overview.tex}

\subsubsection{Discovering Local Communities}\label{sec:local_discovery}

We discover a local community of users with similar interests in a bottom-up manner. Given an entire user set $\mathcal{U}$, we first select $q$ anchor users who are the centers of each local model, denoted by $\mathcal{A} =\{a^{(1)}, \dots, a^{(q)}\}$. (See Section~\ref{sec:anchor} for the selection of anchor users.)

For each anchor user, we discover a set of neighbors by estimating the distance with respect to all other users. As in LLORMA~\cite{LeeKLS13}, we employ the \emph{arccos distance} that is scaled in [0, 1]:
\begin{equation}
s = dist(a^{(j)}, u) = arccos \left(\frac{\mathbf{a}^{(j)} \cdot \mathbf{u}}{\| \mathbf{a}^{(j)} \| \cdot \|\mathbf{u}\|} \right),
\end{equation}
where $\mathbf{a}^{(j)}$ and $\mathbf{u}$ are the embedding vectors for the anchor user $a^{(j)}$ and the user $u$. The embedding vectors are computed by a pre-trained model. (We employ an autoencoder-based model in our experiment, but any other embedding vectors can be used.)

We then apply a smoothing kernel function $K_h(s)$ to adjust the similarity between the users. The kernel is a symmetric uni-modal function, returning a non-negative weight when the distance is within a bandwidth $h$; it covers a broader range with larger $h$ and a narrower range with smaller $h$. (In our experiment, we consider the Epanechnikov kernel function as in~\cite{LeeKLS13}.)

In this paper, we adopt different vector sets for training and inference, to address the issues stated in Section~\ref{sec:motivation} (Figure~\ref{fig:all_data_sim}). Given a set of anchor users $\mathcal{A}$, we build two weight sets $\mathcal{T} =\{\mathbf{t}^{(1)}, \dots, \mathbf{t}^{(q)}\}$ and $\mathcal{W} =\{\mathbf{w}^{(1)}, \dots, \mathbf{w}^{(q)}\}$, where $\mathbf{t}^{(j)} = (t^{(j)}_1, \dots, t^{(j)}_m)$ and $\mathbf{w}^{(j)} = (w^{(j)}_1, \dots, w^{(j)}_m)$ are the weight vectors of $m$ users for the $j$-th anchor user $a^{(j)}$, for training and inference, respectively. They are the vectors of similarities between users and the $j$-th anchor user $a^{(j)}$ with different bandwidths $h_T$ and $h_W$:
\begin{align}
    K_{h_T}(s) & \propto (1 - s^2) \mathbb{1}[s < h_T],
    \label{eq:train_kernel}\\
    K_{h_W}(s) & \propto (1 - s^2) \mathbb{1}[s < h_W],
    \label{eq:aggregate_kernel}
\end{align}
where $\mathbb{1}$ is the indicator function (1 if the condition holds, and 0 otherwise).

It is critical to assign a proper weight $\mathbf{w}^{(j)}$ as only a small number of strongly-tied users in the local community benefit from the prediction of the local model, \ie, $\mathbf{w}^{(j)}$ should be conservative enough to allow only for the core members to use the local model for inference. When noisy predictions of weakly-related local models are involved in aggregation, they may deteriorate the performance. Therefore, our model selectively aggregates local models by preserving coherent sub-communities. (In our experiment, we adjust $h_W$ from $0.2$ to $0.8$, where the average number of users in the local model is approximately $1$--$10\%$.)

\subsubsection{Training Local Models}\label{sec:training}

Each local model is trained on a sub-community with its corresponding weight $\mathbf{t}^{(j)}$. The objective function of learning a local model is formulated as follows:
\begin{equation}
  \argmin_{\theta^{(j)}}\sum_{\mathbf{r}_u \in \mathbf{R}}{t^{(j)}_u \mathcal{L} \left( \mathbf{r}_u, M^{\text{local}}(\mathbf{r}_u; \theta^{(j)}) \right)} + \lambda \Omega(\theta^{(j)}),
  \label{eq:train_local}
\end{equation}
where $\mathbf{r}_u$ is the user vector of $\mathbf{R}$, and $t^{(j)}_u$ is the scalar weight for the user vector $\mathbf{r}_u$. When $t^{(j)}_u$ is zero, the corresponding user $u$ is ignored for training the local model. When $\mathbf{t}^{(j)}$ is all ones, it is equivalent to training the global model.

For the training, we use the bandwidth $h_T$ (usually larger than $h_W$ used for inference) to cover sufficiently many users. As illustrated in Figure~\ref{fig:all_data_sim}, considering more neighbors with a larger bandwidth than $h_W$ yields in capturing correlations among users. When $h_T$ is too small, most $t^{(j)}_u$ becomes zero, leading to an extremely small sub-matrix to train. This makes it difficult to capture hidden local patterns of users, incurring a sub-optimal performance.

\subsubsection{Combining Local Models}\label{sec:aggregation}

As in the existing studies~\cite{LeeKLS13, LeeKLSB16, WangPJSW16, Christakopoulou16, Christakopoulou18}, LOCA aggregates the final predictions from the multiple local models by aggregating their weights, where the weight is proportional to the strength of membership of the target user in the local model. Unlike traditional models, however, LOCA uses a weight set $\mathcal{W}$ that is different from $\mathcal{T}$ used for training. A tighter (smaller) threshold $h_W$ in than $h_T$ in Eq.~\eqref{eq:aggregate_kernel} ensures that the inference relies strictly on similar sets of users.

Also, we train the global model and combine it and multiple local models. The global model is used to learn global correlations among all the users, which is equivalent to assigning equal weights to all the users. As each local model discovers relatively small coherent sub-communities, combining them may not ensure coverage of the entire set of users. To overcome this problem, we employ a global model that is trained by:
\begin{equation}
  \argmin_{\theta^{(g)}}\sum_{\mathbf{r}_u \in \mathbf{R}}{\mathcal{L} \left( \mathbf{r}_u, M^{\text{global}}(\mathbf{r}_u; \theta^{(g)}) \right)} + \lambda \Omega(\theta^{(g)}),
  \label{eq:train_global}
\end{equation}
where $\mathbf{r}_u$ is the user vector of user $u$ in $\mathbf{R}$, and $\theta^{(g)}$ is the parameter for the global model.

Finally, we aggregate the global model and multiple local models using a non-parametric regression method:
\begin{equation}
  \mathbf{\hat{R}} = \alpha M^{\text{global}}(\mathbf{R}; \theta^{(g)}) + (1 - \alpha) \sum_{j=1}^{q}{\mathbf{w}^{(j)} \odot M^{\text{local}}(\mathbf{R}; \theta^{(j)}) \oslash \mathbf{w}},
  \nonumber
  \label{eq:global_local_aggregation}
\end{equation}
where $\mathbf{w}^{(j)}$ is the aggregation weight for the $j$-th local model, $\mathbf{w} = \sum_{j=1}^{q}{\mathbf{w}^{(j)}}$, and $\odot$ and $\oslash$ means the element-wise product and division, respectively. When multiple local models do not cover some of the users, the global model can still compensate for this by considering only the global correlations among the users, \ie, $\alpha = 1 - \mathbb{1}[\exists \mathbf{w}^{(j)} | \mathbf{w}^{(j)} > 0]$.

\subsection{Coverage-based Anchor Selection}
\label{sec:anchor}

To select the anchor users, we adopt a greedy method to maximize the coverage of local models, instead of random selection as in~\cite{LeeKLS13}. We define an unweighted undirected graph $\mathcal{G} = (\mathcal{U}, E)$ of users, where the edges represent the relationships between the users. As the weight $\mathcal{W}$ directly affects the performance of our model, we assign an edge between two users only when the similarity between them is within the kernel bandwidth $k_{h_W}$, \ie, edge $E_{ij} = 1$ if $K_{h_W} (dist (i, j)) > 0$, otherwise $E_{ij} = 0$.

The overall procedure of our anchor selection is as follows. We define a set of anchors $\mathcal{A}$ and a set of covered nodes $\mathcal{C}$, both initialized as empty sets. At the $j$-th step ($j = 1, ..., q$), among the nodes in $\mathcal{U} - \mathcal{A}$, we choose one that is connected to the highest number of uncovered nodes (\ie, nodes not in $\mathcal{C}$). This node becomes the next anchor $a^{(j)}$. We add this new anchor to $\mathcal{A}$, and the newly covered nodes by $a^{(j)}$ to $\mathcal{C}$. We repeat the process until $q$ anchors are chosen. If all the nodes are already covered before having $q$ anchors, we have removed all the nodes from $\mathcal{C}$ and continued the procedure, so that most of the users were covered twice, and so on. (As we usually use a narrow $h_W$, this case does not happen in our experiment.) It is beneficial for improving the performance of our model as the number of local models increases.

The time complexity of the greedy method is $O(q \cdot m^2)$, where the worst case occurs with a densely connected graph. In practice, edges are only sparsely connected ($|E| \approx O(m)$), especially with a narrow kernel bandwidth $K_{h_W}$, leading to an average time complexity of $O(q \cdot m)$. Some of the previous studies~\cite{Christakopoulou16,Christakopoulou18,LeeKLSB16} developed a clustering-based or a distance-based anchor selection. Although they help to maximize the coverage of local models, these methods incur high computational overhead and low performance. By contrast, our coverage-based anchor selection chooses anchor users progressively at a low computational cost.

%% file: Fig2_similarity.tex
\begin{figure}
\includegraphics[width=0.7\linewidth]{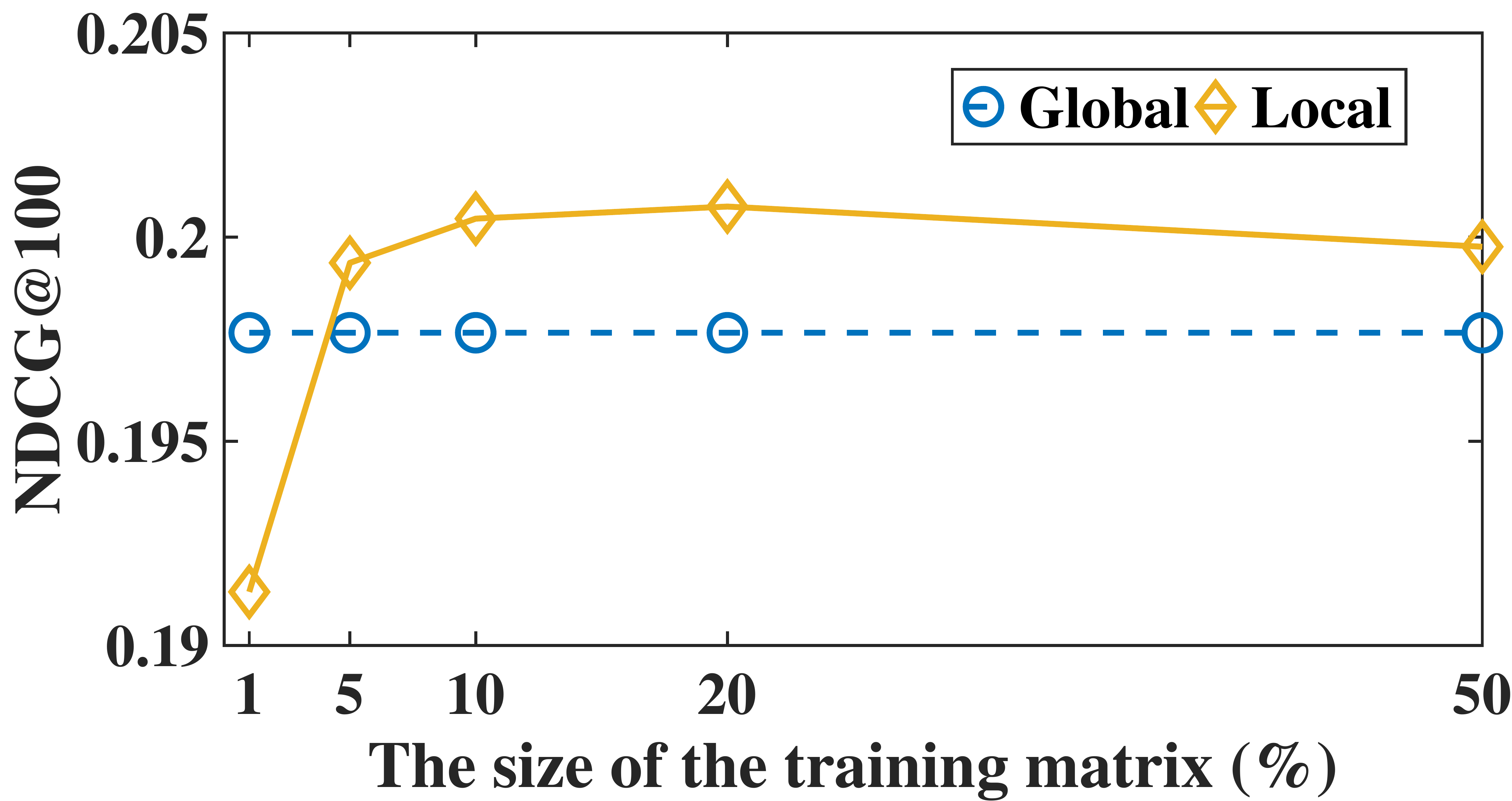}
\caption{A motivating example showing superior performance of local models trained on a larger (top 5--50\%) neighborhood than that for inference (top 1\%). (This result is an average of 100 local models centered on a random anchor user, trained on MovieLens 10M.)
}\label{fig:all_data_sim}
\vskip -0.15in
\end{figure}

%% file: Fig3_overview.tex
\begin{figure}[t]
\centering
\includegraphics[width=1.08\linewidth]{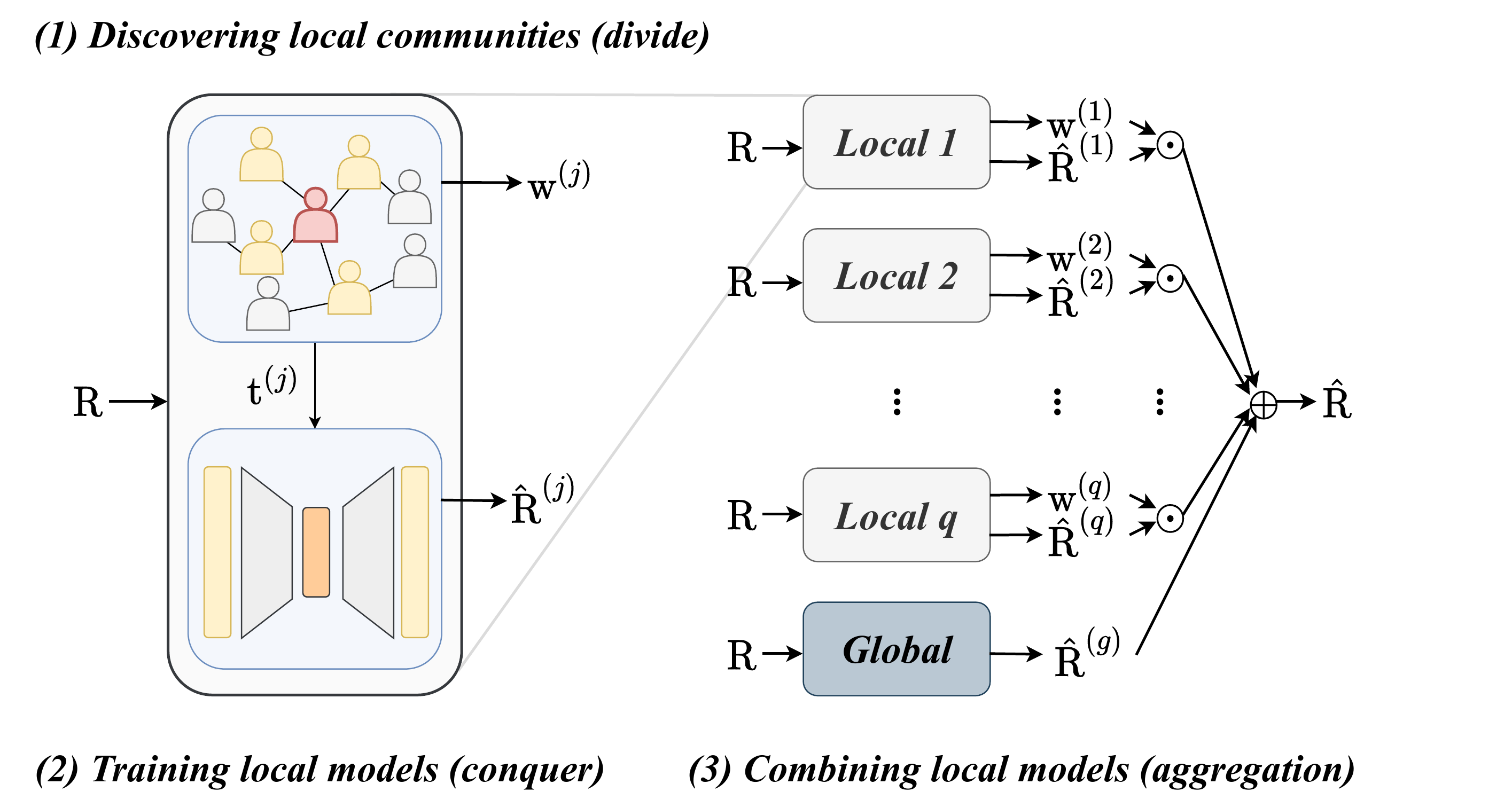}
\vskip -0.1in
\caption{Model architecture of LOCA. We used $\mathcal{T}$ and $\mathcal{W}$ to train and infer local models. Both the weights are calculated on the basis of each user's similarity with an anchor user.}\label{fig:Overview}
\vskip -0.15in
\end{figure}

%% file: sec-experiments.tex
\section{Experimental Setup}\label{sec:setup}

\subsection{Datasets and Baselines}\label{sec:dataset}

\input{Tab1_dataset.tex}

Table~\ref{tab:statistics} summarizes the statistics of the datasets used in this paper. We used five public benchmark datasets: MovieLens 10M (ML10M), MovieLens 20M (ML20M), Amazon Digital Music (AMusic), Amazon Video Games (AGames), and Yelp 2015 (Yelp). They were selected to span various data sparsity levels and sizes of the rating matrix. We converted all the explicit ratings to binary values and removed the users with ratings less than 10. Detailed pre-processing and all source codes can be found on our website\footnote{https://github.com/jin530/LOCA}.


We adopted MutlVAE~\cite{LiangKHJ18} and EASE$^{R}$~\cite{Steck19} as the base models of LOCA, \ie, LOCA$_{VAE}$ and LOCA$_{EASE}$. We compared our models with the global models such as WMF~\cite{HuKV08}, SLIM~\cite{NingK11}, CDAE~\cite{WuDZE16}, MultVAE~\cite{LiangKHJ18}, EASE$^{R}$~\cite{Steck19}, and RecVAE~\cite{ShenbinATMN20}, the latter two being the state-of-the-art models exhibiting the best performance on large-scale datasets. We compare our models with the local models such as LLORMA~\cite{LeeKLS13,LeeKLSB16}, and sGLSVD~\cite{Christakopoulou18}, where MF~\cite{Koren08} and SVD~\cite{CremonesiKT10} were the base models, respectively.

There are more neural recommender models with different architectures such as NeuMF~\cite{HeLZNHC17}, NAIS~\cite{HeHSLJC18}, and NGCF~\cite{Wang0WFC19}. As reported by \citet{LiangKHJ18}, MultVAE~\cite{LiangKHJ18} outperforms NeuMF~\cite{HeLZNHC17}; hence, we excluded NeuMF from our baselines. Moreover, as LOCA is built upon the autoencoder-based models, we mainly employed them as the competing models.

\subsection{Evaluation Protocol and Metrics}
\label{sec:metrics}

We adopted timestamp-based \emph{leave-k-out} evaluation method~\cite{HeLZNHC17,XueDZHC17,HeDWTTC18}. For each user, we held-out the last $k$ feedback for evaluation and used the rest for training. In the existing work, $k$ was usually set to 1. However, we observed that when $k$ was too small, the variance in the results was large, \ie, the local models were subject to overfitting (potentially more than a global model as they were trained on smaller subsets). Thus, local models tend to achieve much higher accuracy when $k = 1$. For a fair comparison, we thus set $k = 5$, indicating more stable accuracy.

For testing, we regarded all unrated items as candidates. Some previous studies~\cite{HeLZNHC17, XueDZHC17, HeDWTTC18} employed a sampling-based evaluation that randomly chose 100 items out of all unrated items for efficient evaluation. As our evaluation considered all unrated items as test items, we believe that it provides more realistic use cases.

We employed two metrics, \emph{Recall@N} and \emph{Normalized Discounted Cumulative Gain (NDCG)}, for evaluating the accuracy of the \topN\ recommendation~\cite{HeLZNHC17,XueDZHC17,HeDWTTC18}. Recall@$N$ checks the number of test items included in the \topN\ list. The score is defined as $\frac{\sum_{i=1}^N{rel_i}}{k}$, where $i$ is the position in the list, $k$ is the number of test items and $rel_i \in \{0, 1\}$ indicates whether the $i$-th item is relevant to the user or not. NDCG@$N$ takes into account the order of retrieved items in the list. DCG@$N$ is defined as $\sum_{i=1}^N \frac{2^{rel_i} - 1}{log_2 (i+1)}$. NDCG is the ratio of DCG to the maximum possible DCG for that user, which occurs when the recommended items are presented in decreasing order of user preference. We chose $N = \{50, 100\}$ for the \topN\ recommendation.

\subsection{Implementation Details}\label{sec:metrics}

For all the gradient-learning-based models, we initialized the parameters from a normal distribution $\mathcal{N}(0, 0.01)$ and trained them using an Adam optimizer~\cite{KingmaB14}. By default, the batch size was 512. We conducted the grid search over the hyperparameters of all the models, and tuned the learning rate among \{0.1, 0.01, 0.001\} and the $L_2$-regularization term among \{0.001, 0.01, 0.1\}. As an early stopping condition, we used 50 epochs of patience.

For WMF~\cite{HuKV08}, LLORMA~\cite{LeeKLS13, LeeKLSB16}, and sGLSVD~\cite{Christakopoulou18}, we tuned the $\lambda_{u}$ and $\lambda_{i}$ among \{1, 10, 100\} and chose the $\alpha$ among \{2, 5, 10\}. For AE-based models, we set the dimension of the latent representation to 200 and tuned the dropout ratio among \{0, 0.2, 0.5\}. For MultVAE~\cite{LiangKHJ18}, we did not apply the $L_2$-norm regularization and used either 0 or 1 additional MLP layers, as discussed in~\cite{LiangKHJ18}. For EASE$^{R}$~\cite{Steck19}, we tuned the $L_2$-norm regularization parameter $\lambda$ among \{1, 10, 100, 1000\}.  For RecVAE~\cite{ShenbinATMN20}, we tuned the $\beta$ among \{0.1, 0.01, 0.001\}.  For LLORMA~\cite{LeeKLS13,LeeKLSB16}, we set the number of local models as 100 and tuned the kernel bandwidth $h$ = \{0.8, 1.0, 2.0, 4.0, 5.0\}. For sGLSVD~\cite{Christakopoulou18}, we tuned the number of local models using the clustering algorithm among \{2, 3, 5, 10, 20, 50, 100\}, as discussed in~\cite{Christakopoulou16}. For LOCA, we tuned the training kernel bandwidth $h_T$ in $[0.8, 2.0]$ and aggregating bandwidth $h_W$ in $[0.2, 0.8]$.

We conducted all the experiments on a desktop with 2 Nvidia TITAN RTX, 256 GB memory, and 2 Intel Xeon Processor E5-2695 v4 (2.10 GHz, 45M cache). We used a public package\footnote{https://github.com/KarypisLab/SLIM} for SLIM. In addition, we implemented all the gradient-learning-based models using PyTorch\footnote{https://pytorch.org/}. 

\input{Tab2_results.tex}

%% file: Tab1_dataset.tex
\begin{table}[t]
\small
\begin{center}
\caption{Statistics of five public benchmark datasets. Concentration is the ratio of the ratings of the top 5\% of the most popular items.}
\label{tab:statistics}
\begin{tabular}{l|rrrrr}
\toprule
Dataset & \#Users & \#Items & \#Ratings & Sparsity & Concentration \\
\midrule
ML10M & 69,878 & 10,677 & 10,000,054 & 98.66\% & 48.04\% \\
ML20M & 138,493 & 26,744 & 20,000,263 & 99.46\% & 66.43\% \\
AMusic & 4,964 & 11,797 & 97,439 & 99.83\% & 14.93\% \\
AGames & 13,063 & 17,408 & 236,415 & 99.90\% & 16.40\%\\
Yelp & 25,677 & 25,815 & 731,671 & 99.89\% & 22.78\% \\
\bottomrule
\end{tabular}
\end{center}
\end{table}

%% file: Tab2_results.tex
\begin{table*}[t] \small
\caption{Comparison of the performance of LOCA (LOCA$_{VAE}$ and LOCA$_{EASE}$) and competing models. Gain indicates the accuracy improvement of LOCA is over the best baseline model. The best model is marked in \textbf{bold} and the best baseline is \underline{underlined}.}
\label{tab:overallresult}
\begin{center}
\begin{tabular}{c|c|cccccccccc|c}
\toprule
Dataset                & Metric    & WMF        & SLIM   & CDAE   & MultVAE & EASE$^{R}$    & RecVAE   & LLORMA               & sGLSVD & LOCA$_{VAE}$    & LOCA$_{EASE}$  & Gain   \\ \midrule
\multirow{4}{*}{ML10M}  & Recall@50  & 0.3035           & 0.3069 & 0.3222 & 0.3141              & 0.3258             & 0.3168       & \underline{0.3282}    & 0.3131 & \textbf{0.3367} & 0.3335 & 2.59\% \\
                        & Recall@100 & 0.4333           & 0.4381 & 0.4685 & 0.4653              & 0.4648             & \underline{0.4705} & 0.4692           & 0.4468 & \textbf{0.4865} & 0.4798 & 3.40\% \\
                        & NDCG@50    & 0.1492           & 0.1544 & 0.1580 & 0.1529              & 0.1617            & 0.1544            & \underline{0.1655} & 0.1586 & \textbf{0.1661} & 0.1647 & 0.36\% \\
                        & NDCG@100   & 0.1848           & 0.1904 & 0.1982 & 0.1945              & 0.2000            & 0.1966        & \underline{0.2042}                & 0.1953 & \textbf{0.2073} & 0.2049 & 1.52\% \\ \hline
\multirow{4}{*}{ML20M}  & Recall@50  & 0.2934           & 0.2912 & 0.2958 & 0.3003              & \underline{0.3132}& 0.2975        & 0.2291                & 0.3047 & 0.3032           & \textbf{0.3258} & 4.02\% \\
                        & Recall@100 & 0.4277           & 0.4169 & 0.4324 & 0.4397              & \underline{0.4468}& 0.4417        & 0.3355                & 0.4342 & 0.4419           & \textbf{0.4654} & 4.16\% \\
                        & NDCG@50    & 0.1443           & 0.1486 & 0.1469 & 0.1477              & \underline{0.1581}& 0.1461        & 0.1153                & 0.1564 & 0.1503           & \textbf{0.1641} & 3.80\% \\
                        & NDCG@100   & 0.1812           & 0.1831 & 0.1844 & 0.1860              & \underline{0.1948}& 0.1857        & 0.1446                & 0.1919 & 0.1884           & \textbf{0.2024} & 3.90\% \\ \hline
\multirow{4}{*}{AMusic} & Recall@50  & 0.1118           & 0.0860 & 0.1005 & \underline{0.1177} & 0.1129             & 0.1050          & 0.0925               & 0.0899 & \textbf{0.1235} & 0.1138 & 4.93\% \\
                        & Recall@100 & 0.1555           & 0.1234 & 0.1456 & \underline{0.1689} & 0.1506             & 0.1516         & 0.1366                & 0.1257 & \textbf{0.1764} & 0.1539 & 4.44\% \\
                        & NDCG@50    & 0.0663           & 0.0553 & 0.0588 & 0.0681              & \underline{0.0717} & 0.0582         & 0.0517              & 0.0515 & \textbf{0.0748} & 0.0717 & 4.32\% \\
                        & NDCG@100   & 0.0783           & 0.0656 & 0.0712 & 0.0822 & \underline{0.0821}              & 0.0710           & 0.0638             & 0.0613 & \textbf{0.0893} & 0.0826 & 8.64\% \\ \hline
\multirow{4}{*}{AGames} & Recall@50  & \underline{0.1429} & 0.1182 & 0.1240 & 0.1402            & 0.1394             & 0.1300            & 0.0806            & 0.1167 & \textbf{0.1473} & 0.1400 & 3.08\% \\
                        & Recall@100 & 0.2012           & 0.1631 & 0.1825 & \underline{0.2081} & 0.1913               & 0.1920          & 0.1223             & 0.1669 & \textbf{0.2147} & 0.1947 & 3.17\% \\
                        & NDCG@50    & 0.0763           & 0.0666 & 0.0647 & 0.0733              & \underline{0.0773} & 0.0679         & 0.0425              & 0.0638 & \textbf{0.0781} & 0.0772 & 1.03\% \\
                        & NDCG@100   & \underline{0.0923} & 0.0789 & 0.0808 & 0.0920            & 0.0915               & 0.0849         & 0.0539             & 0.0777 & \textbf{0.0966} & 0.0922 & 4.66\% \\ \hline
\multirow{4}{*}{Yelp}   & Recall@50  & 0.1483            & 0.1222 & 0.1382 & \underline{0.1499} & 0.1459                & 0.1480        & 0.0651               & 0.1291 & \textbf{0.1560} & 0.1472 & 4.07\% \\
                        & Recall@100 & 0.2247            & 0.1804 & 0.2094 & \underline{0.2276} & 0.2187                & 0.2262            & 0.1013             & 0.1965 & \textbf{0.2354} & 0.2205 & 3.43\% \\
                        & NDCG@50    & \underline{0.0774} & 0.0649 & 0.0725 & 0.0769            & 0.0772               & 0.0761            & 0.0330             & 0.0672 & \textbf{0.0812} & 0.0780 & 4.91\% \\
                        & NDCG@100   & \underline{0.0984} & 0.0809 & 0.0920 & 0.0982            & 0.0972                & 0.0975         & 0.0429                 & 0.0857 & \textbf{0.1030} & 0.0981 & 4.67\% \\ \bottomrule
\end{tabular}
\end{center}
\end{table*}

%% file: sec-results.tex
\section{Experimental Results}
\label{sec:results}

In this section, we compare the accuracies of LOCA and the competing models. Through the extensive experiments, we provide the following meaningful insights:
\begin{itemize}[leftmargin=5.5mm]
    \item $[$Sec.~\ref{sec:exp_overall}$]$ LOCA$_{VAE}$ and LOCA$_{EASE}$ achieve state-of-the-art results with an improvement of up to 4.70\% (Recall) and 7.95\% (NDCG) over the existing models, irrespective of whether they are global or local models.
    
    \item $[$Sec.~\ref{sec:exp_localcount}$]$ As the number of local models increases, LOCA$_{VAE}$ shows better performance. With up to 300 local models, it achieves up to 8.63\% improvement gain in NDCG@100 as compared to MultVAE~\cite{LiangKHJ18}.
    
    \item $[$Sec.~\ref{sec:exp_anchor}$]$ The coverage-based anchor selection in LOCA$_{VAE}$ outperforms the other anchor selection methods (an improvement of up to 1.70\% in NDCG@100 over the random selection). 
    
    \item $[$Sec.~\ref{sec:exp_backbone}$]$ LOCA can be integrated with other AE-based models, and the variant of LOCA also consistently outperforms the base models (an improvement of up to 5.39\% in NDCG@100 relative to the global model).
\end{itemize}

\input{Fig4_num_ratings.tex}

\subsection{Effectiveness of LOCA}
\label{sec:exp_overall}

Table~\ref{tab:overallresult} reports the accuracies of LOCA and the competing models on five datasets. The key observations are: (i) LOCA shows better accuracy than the competing models across all datasets. (ii) LOCA consistently outperforms its base models. (iii) The state-of-the-art AE-based models, such as MultVAE and RecVAE, tend to be better than the existing local models such as LLORMA and sGLSVD, implying that neural recommender models can capture global patterns effectively. (iv) Lastly, it is interesting to observe that EASE-based models (EASE$^R$ and LOCA$_{EASE}$) are relatively stronger on larger and denser datasets, such as MovieLens 10M and 20M, whereas VAE-based models (MultVAE and LOCA$_{VAE}$) are stronger on sparser datasets such as AMusics, AGames, or Yelp. Therefore, We surmise that each base model has its own strength and weakness, and putting into the LOCA framework still improves overall performance.

Figure~\ref{fig:breakdown_ratings} depicts the breakdown results on two datasets, according to the number of ratings per user. LOCA$_{VAE}$ shows a consistent improvement in performance over all intervals. This is to be expected as LOCA$_{VAE}$ handles local correlations of users/items, whereas the base model faces challenges in capturing local correlations. Even when the user-item interaction is scarce, LOCA$_{VAE}$ achieves better performance than the base model as it captures the locality of small sub-communities, \eg, long-tail users/items.

\subsection{Effect of Number of Local Models}
\label{sec:exp_localcount}

Figure~\ref{fig:num_model} depicts the extent to which the number of local models affects the performance of LOCA$_{VAE}$. LLORMA$_{VAE}$ builds and aggregates local models based on LLORMA~\cite{LeeKLS13, LeeKLSB16}; however, it adopts MultVAE~\cite{LiangKHJ18} as its base model instead of MF. It was observed that the accuracy of LOCA$_{VAE}$ improved consistently with an increase in the number of local models. For the motivating question that we posed in Section~\ref{sec:motivation}, \emph{is it better to use different weights for training and aggregating local models?}, our experimental results imply that the answer is yes. The simple ensemble model that assigns the same weight values for all local models and LLORMA$_{VAE}$ using the same weights for training and aggregating local models show a limited improvement, converging to the best NDCG@100 with 50 local models. By contrast, LOCA$_{VAE}$ improves beyond this level by taking advantage of up to 300 newly added local models. From this observation, we conclude that LOCA$_{VAE}$ is more effective in capturing local sub-communities. One might wonder whether this conclusion would hold if we had as many local models the number of users. Developing more local models does not impair the accuracy, but the performance gains diminish.

\input{Fig5_num_local.tex}

\subsection{Effect of Anchor User Selection}
\label{sec:exp_anchor}

Table~\ref{tab:anchorselection} compares the different anchor selection strategies, including random, $k$-means and farthest\footnote{Originally named as ``Coverage'' in \cite{LeeKLSB16}. To avoid confusion, we rename this as ``farthest'', as this method maximizes the distance across anchor users.}~\cite{LeeKLSB16}. Although the difference in performance between the coverage-based selection and the other methods is not significant, the former consistently outperforms other selection methods. As reported in~\cite{LeeKLSB16}, the cluster-based selection shows a better performance than the random and farthest selection methods; however, it is less effective than the coverage-based selection. We conclude that the coverage-based selection effectively increases the number of local models, thereby obtaining consistent gains of LOCA.

\subsection{Effect of Various Base Models}
\label{sec:exp_backbone}

To investigate the model-agnostic property of LOCA, we employ different base models, \ie, DAE~\cite{WuDZE16} and MultVAE~\cite{LiangKHJ18}. 
Compared to the base models, Figure~\ref{fig:backbone} shows that both LOCA$_{DAE}$ and LOCA$_{VAE}$ consistently outperform their global counterparts. Moreover, as the number of local models increases, we observe consistent performance gains. These results imply that a locally low-rank prior is generally valid regardless of specific base models. Therefore, we conclude that LOCA can be easily applied to other base models.

\input{Fig6_backbone.tex}
\input{Fig78_hyperparameters.tex}

\subsection{Effect of Hyperparameters}
\label{sec:exp_hyperparam}

Figure~\ref{fig:train_h} depicts the effect of the kernel bandwidth $h_T$ in training local models. When the bandwidth $h_T$ is $1.0$--$4.0$, LOCA achieves the best accuracy, where the coverage of each local model is approximately $64.5$--$100\%$. When the bandwidth is too low, i.e., $h_T$ is $0.5$, the accuracy decreases as the number of local models increases. This implies that the coverage of the local model must be adequate to capture local correlations for users.

Figure~\ref{fig:aggregate_h} depicts the effect of kernel bandwidth $h_W$ in aggregating local models. When the bandwidth is $0.4$--$0.8$, the performance is the best, where the coverage of 300 local models is approximately  $25.7$--$100$\%. When the bandwidth is low, \ie, $h_W$ is $0.2$--$0.4$, the accuracy for specific users covered by the local model increases, but this has little effect on the accuracy for the entire set of users. Coupling this with our observation in Figure~\ref{fig:train_h}, we surmise that one reason for LOCA to outperforming LLORMA and its variants is its flexibility in using different weights for training and aggregating local models.

\input{Tab3_anchor.tex}

\subsection{Illustration of Local Models}
\label{sec:visualization}


Table~\ref{tab:casestudy} shows the differences between local models. For the demonstration, we chose a target user who mostly rated the movies for two genres, Sci-Fi and Horror, in the ML10M dataset. The global model attempts to capture both tastes without identifying the different genres. By contrast, the local models in LOCA$_{VAE}$ distinguish different local patterns of user preferences; the first (Local 70) mostly recommends Sci-Fi movies, whereas the second (Local 179) mainly suggests horror movies. Some of the recommendations cover the ground-truth (red-box). We conclude that our local models capture meaningful local patterns, and LOCA$_{VAE}$ successfully utilizes sub-communities consisting of different local patterns.

\subsection{Training Time}\label{sec:exp_training}

The training time of a single local model in LOCA$_{VAE}$ is generally less than that of a global model, as it tackles a smaller problem. Further, local models are independent of each other; therefore, we can train them in parallel. Given sufficient computation power, LOCA with hundreds of local models can be trained within the time taken to train a single local model.

Table~\ref{tab:traintime} presents the actual training times measured for a global model (MultVAE~\cite{LiangKHJ18}) and LOCA for several benchmark datasets. First, the time taken to train a single local model is approximately 2--6 times less than that for MultVAE~\cite{LiangKHJ18}. We also observe that the training time for 1, 2, and 5 local models is not significantly different, as we utilize 8x-parallelism with a multi-core. For 10 or more local models, the training time increases linearly; the increase in training time within the effective parallelism from 1 to 5 is mainly due to the variance in the size of the local model. The datasets with higher concentration shown in Table~\ref{tab:statistics}, \eg, MovieLens, tend to have higher variance in the size of the local model, leading to a longer waiting time for the largest local model.

\input{Tab4_case_study}

%% file: Fig4_num_ratings.tex
\begin{figure}
\centering
\begin{tabular}{cc}
\includegraphics[width=0.220\textwidth]{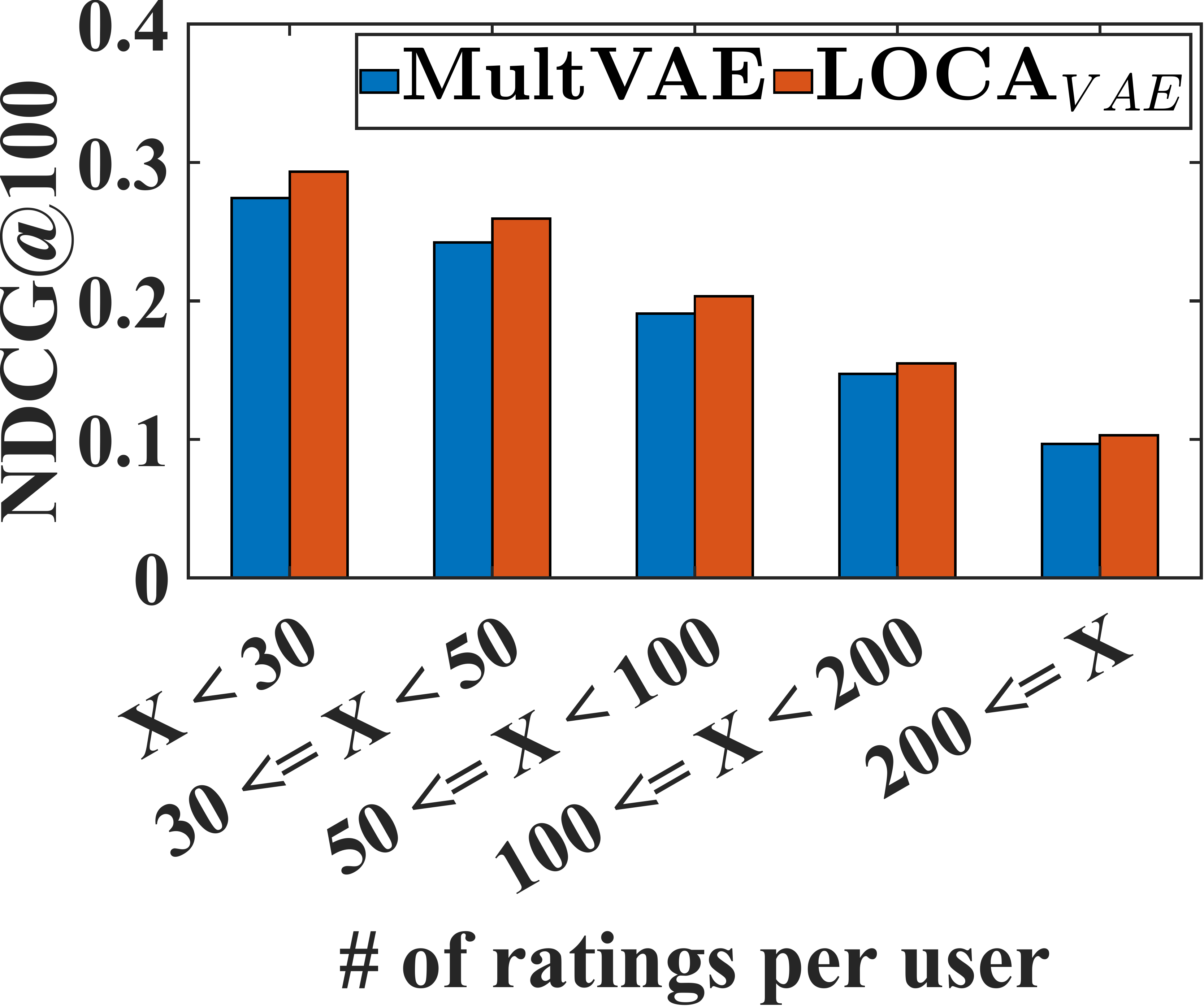} &
\includegraphics[width=0.220\textwidth]{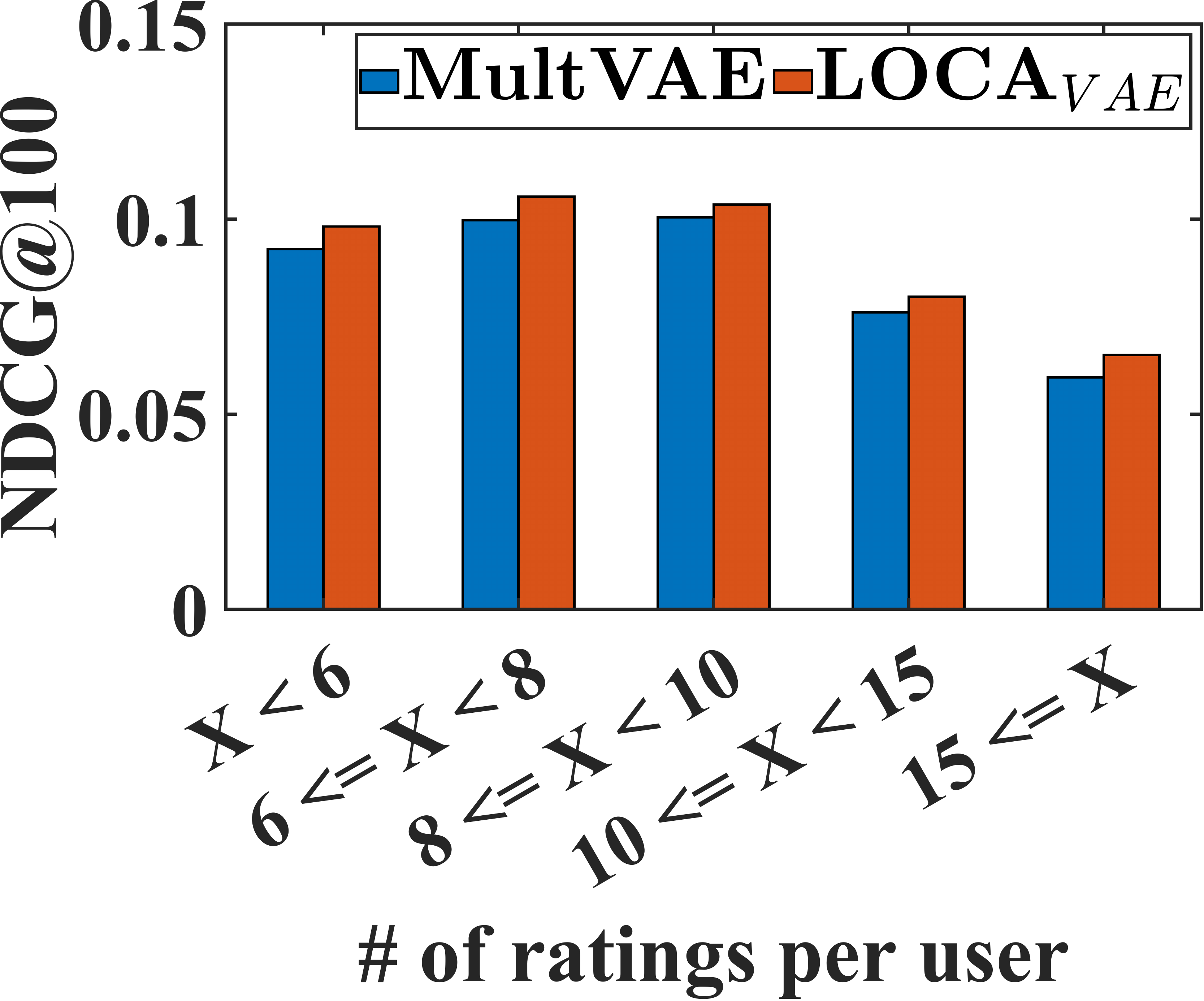} \\
(a) ML10M & (b) AMusic \\
\end{tabular}
\vskip -0.1in
\caption{NDCG@100 with the different number of ratings per user ($X$) for two benchmark datasets.}
\label{fig:breakdown_ratings}
\vskip -0.15in
\end{figure}

%% file: Fig5_num_local.tex
\begin{figure}
\centering
\begin{tabular}{cc}
\includegraphics[width=0.220\textwidth]{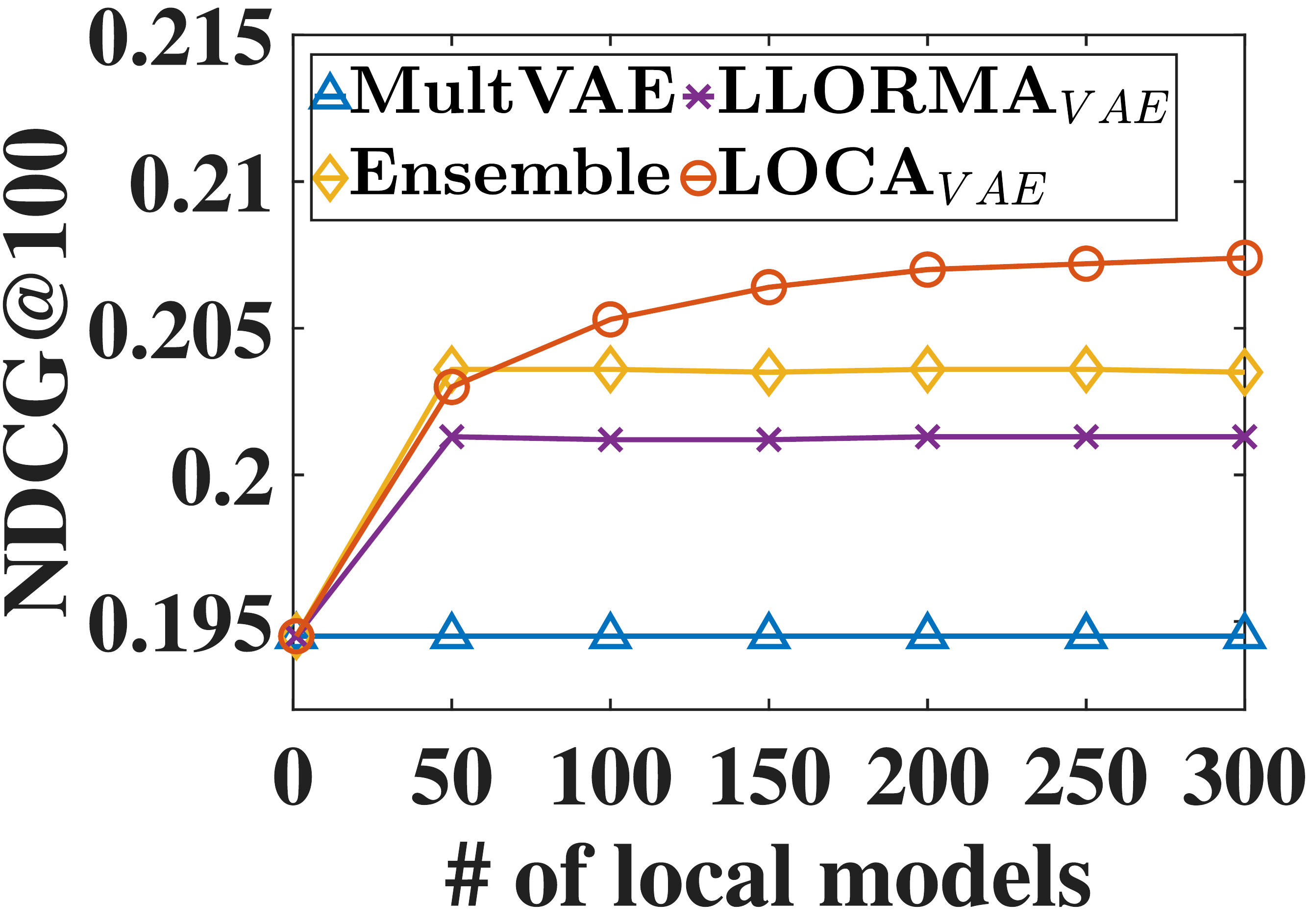} &
\includegraphics[width=0.220\textwidth]{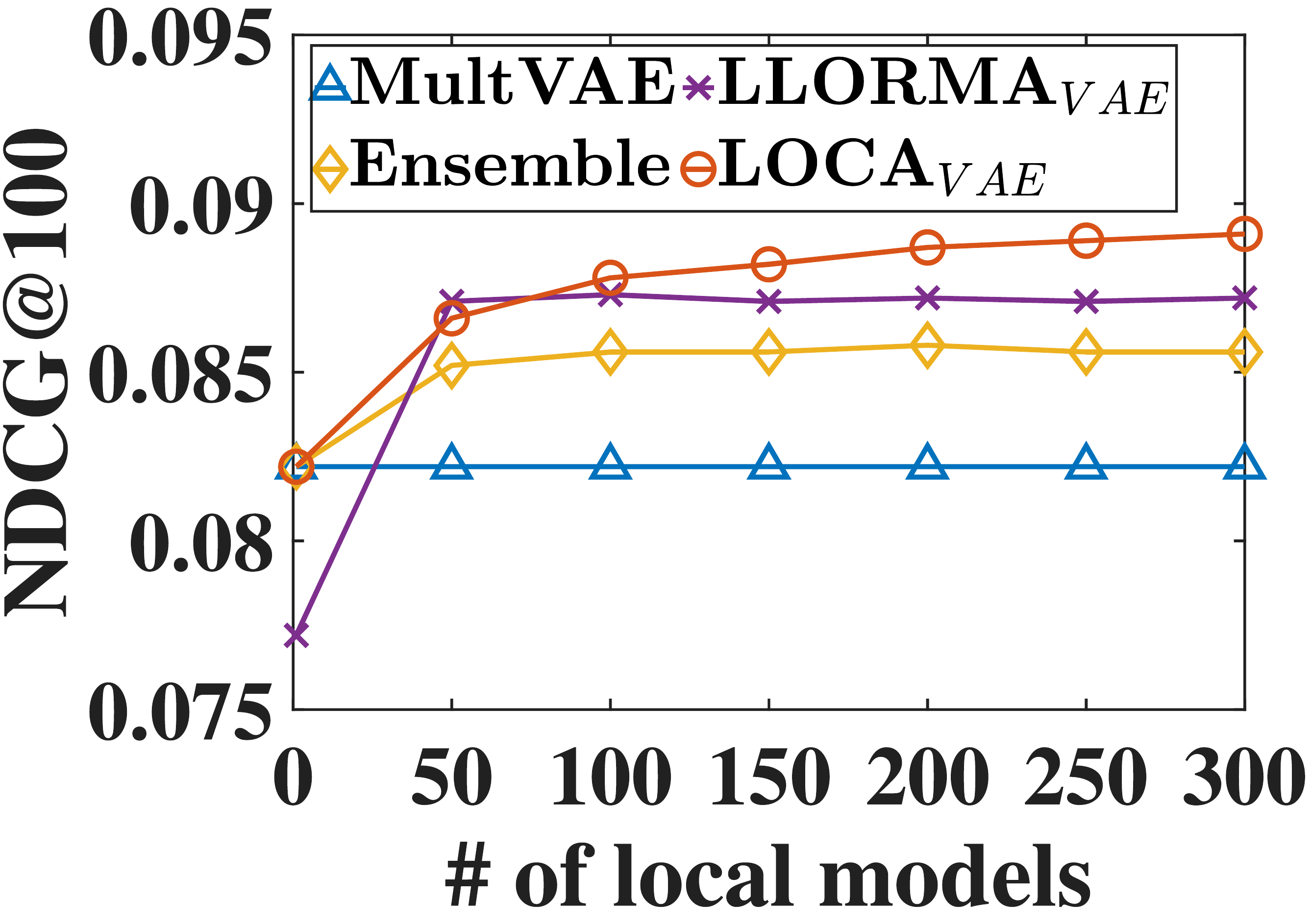} \\
(a) ML10M & (b) AMusic \\
\end{tabular}
\vskip -0.1in
\caption{Comparison of NDCG@100 of MultVAE~\cite{LiangKHJ18}, ensemble, LLORMA$_{VAE}$, and LOCA$_{VAE}$ for two benchmark datasets.}\label{fig:num_model}
\vskip -0.2in
\end{figure}

%% file: Fig6_backbone.tex

\begin{figure}[t!]
\centering
\begin{tabular}{cc}
\includegraphics[width=0.22\textwidth]{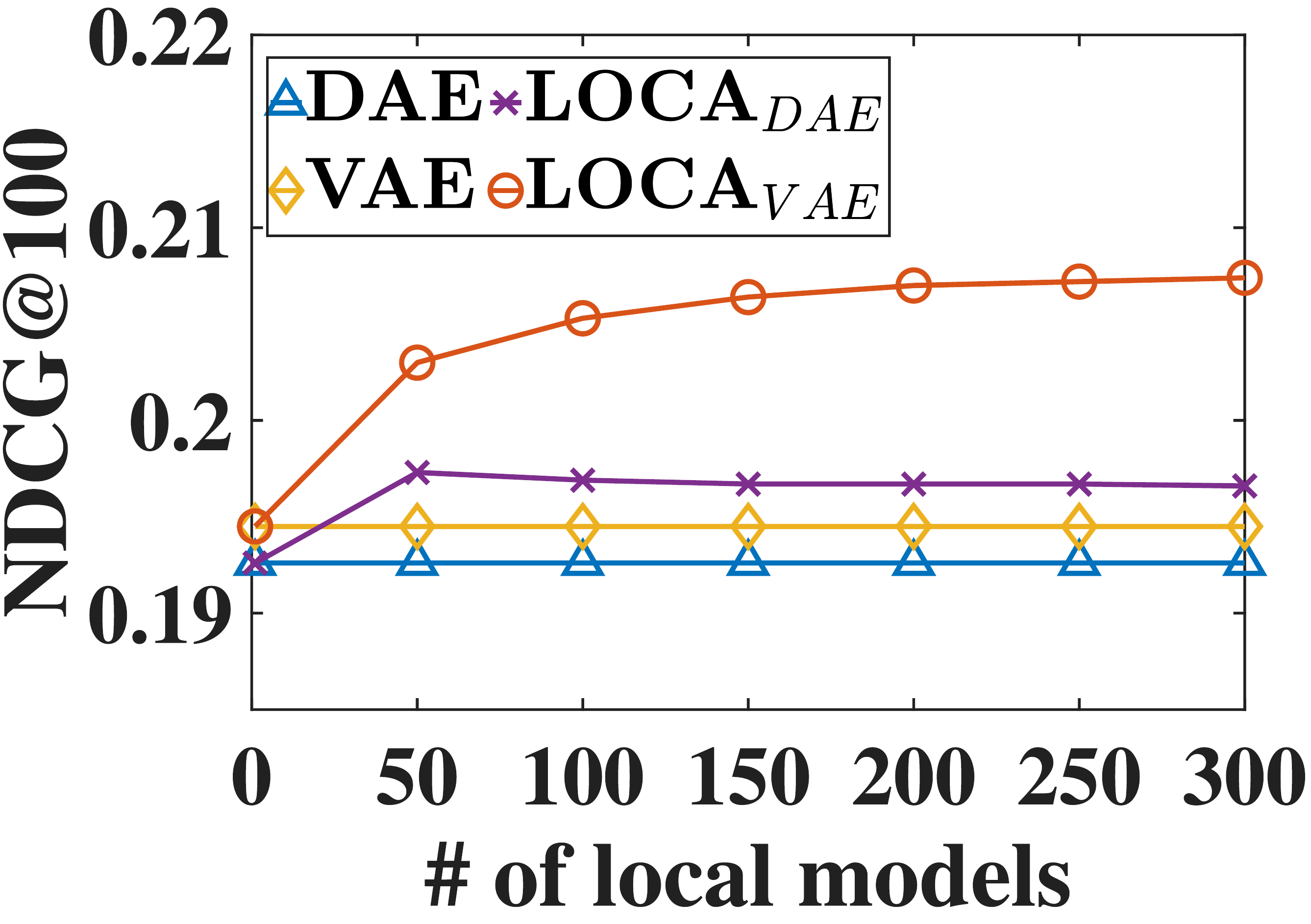} &
\includegraphics[width=0.22\textwidth]{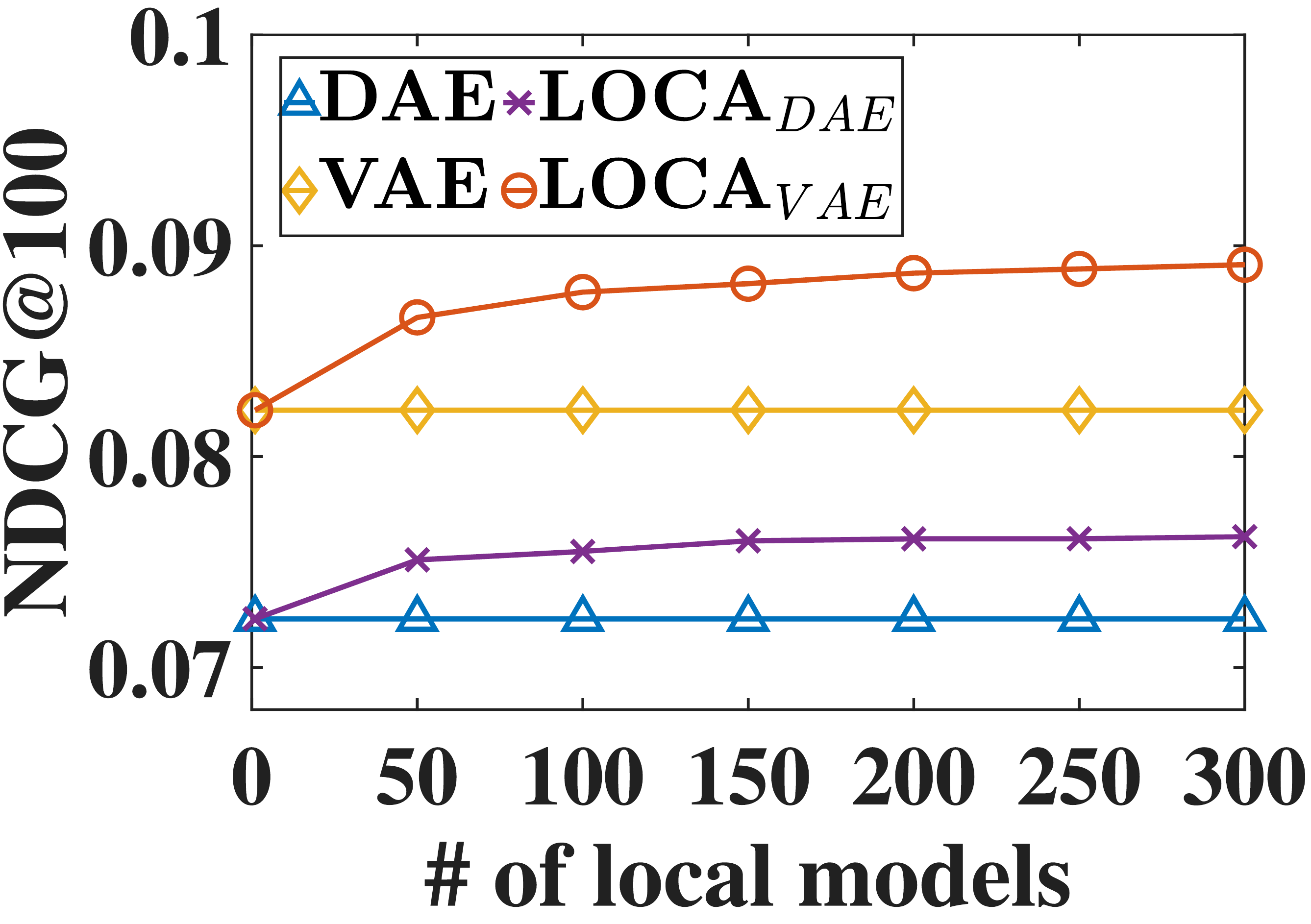} \\
(a) ML10M & (b) AMusic \\
\end{tabular}
\vskip -0.1in
\caption{NDCG@100 of LOCA with various base models (\ie, DAE~\cite{WuDZE16} and MultVAE~\cite{LiangKHJ18}).}\label{fig:backbone}
\vskip -0.15in
\end{figure}

%% file: Fig78_hyperparameters.tex

\begin{figure}[t]
\centering
\begin{tabular}{cc}
\includegraphics[width=0.22\textwidth]{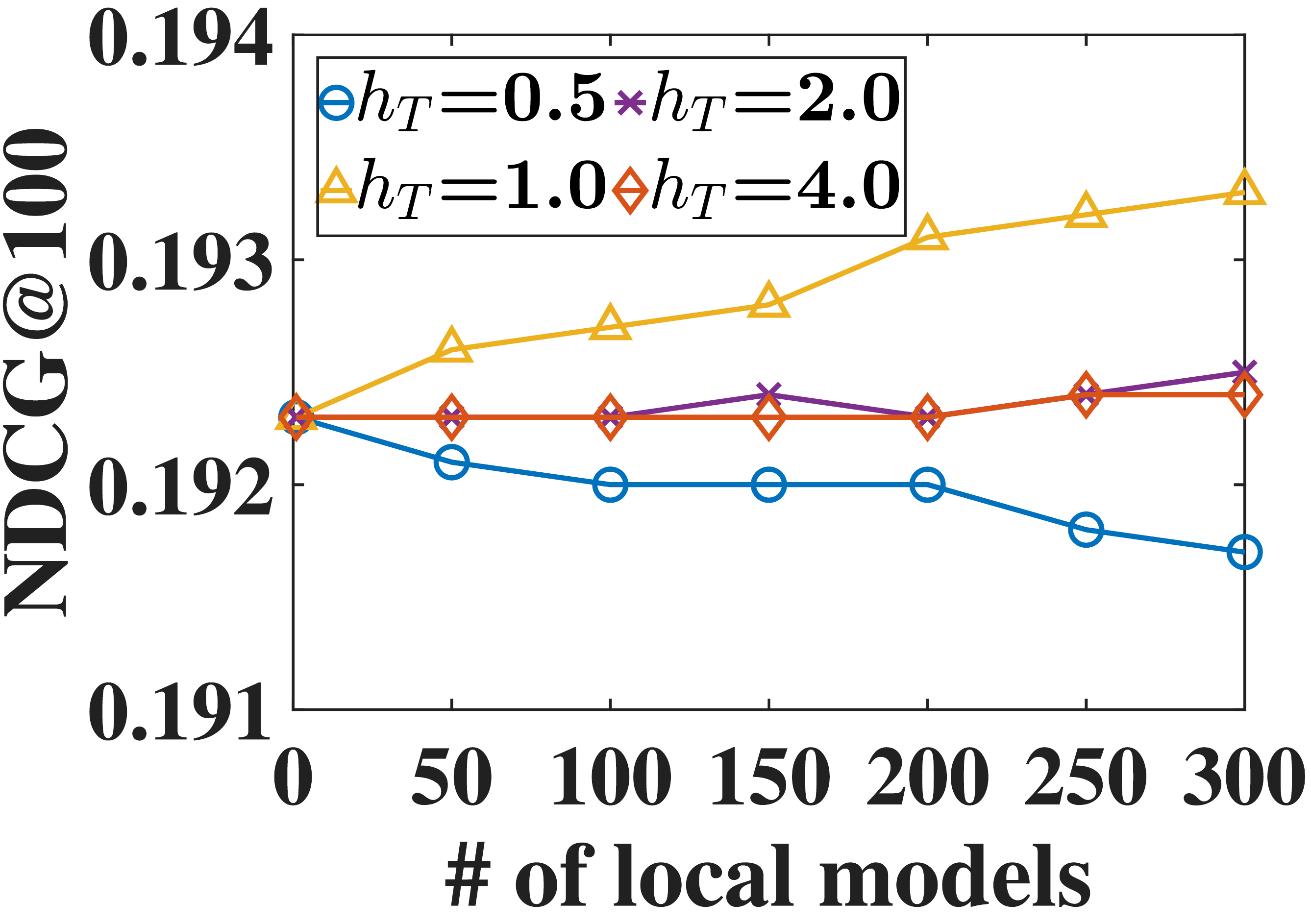} &
\includegraphics[width=0.22\textwidth]{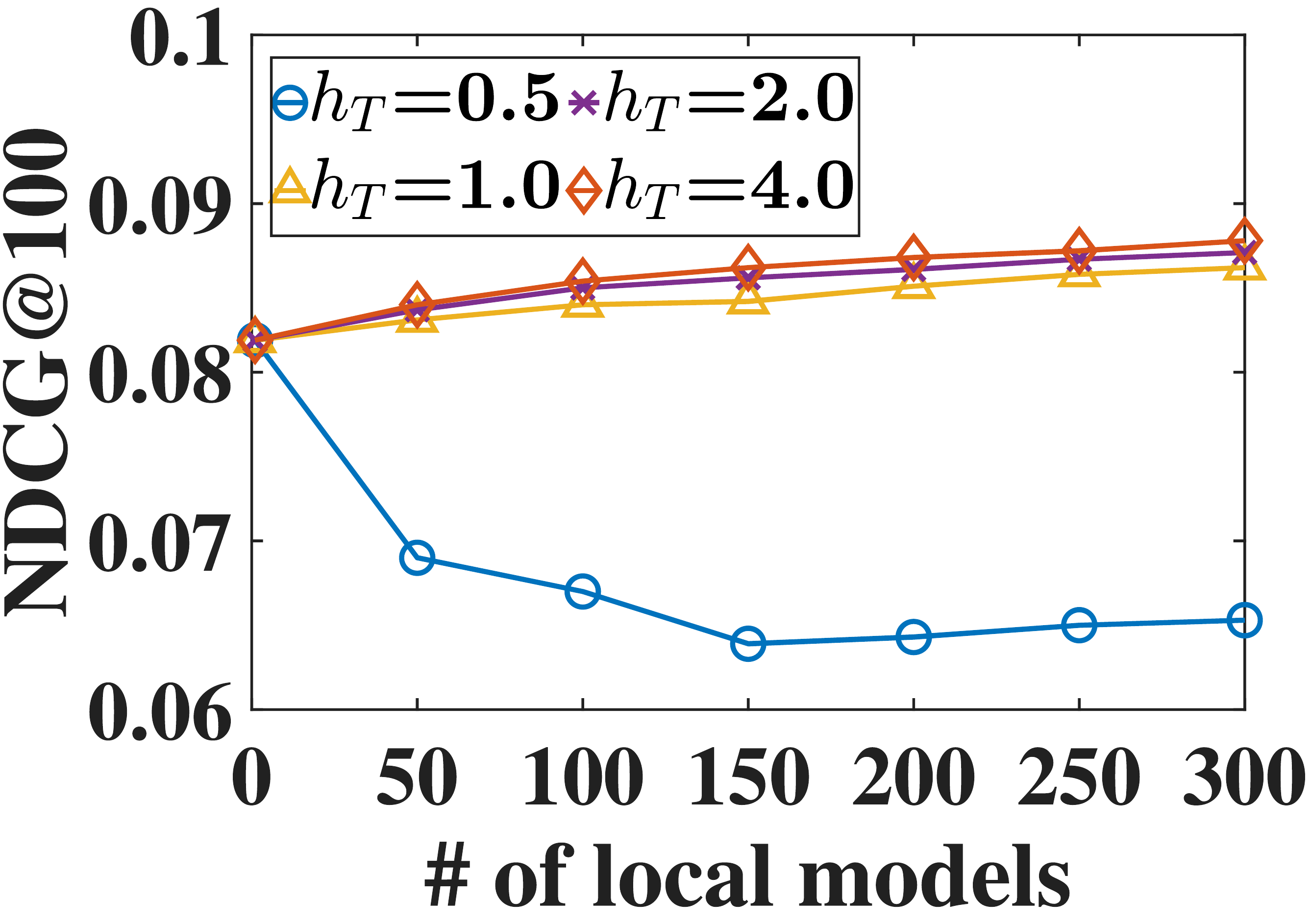} \\
(a) ML10M & (b) AMusic \\
\end{tabular}
\vskip -0.1in
\caption{NDCG@100 vs. kernel bandwidth $h_T$ in training local models. ($h_W=0.4$)}
\label{fig:train_h}
\vskip -0.15in
\end{figure}

\begin{figure}[t]
\centering
\begin{tabular}{cc}
\includegraphics[width=0.22\textwidth]{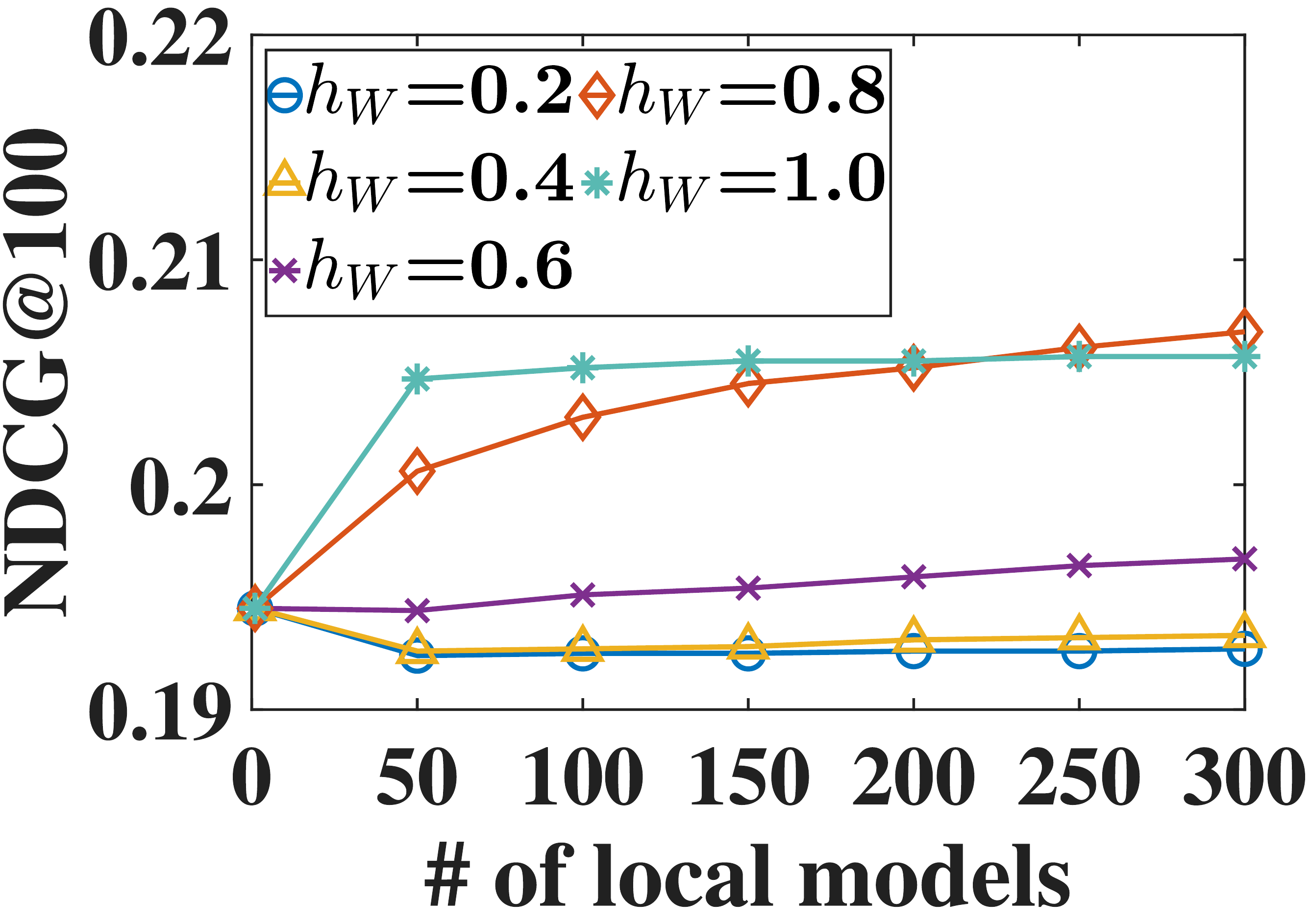} &
\includegraphics[width=0.22\textwidth]{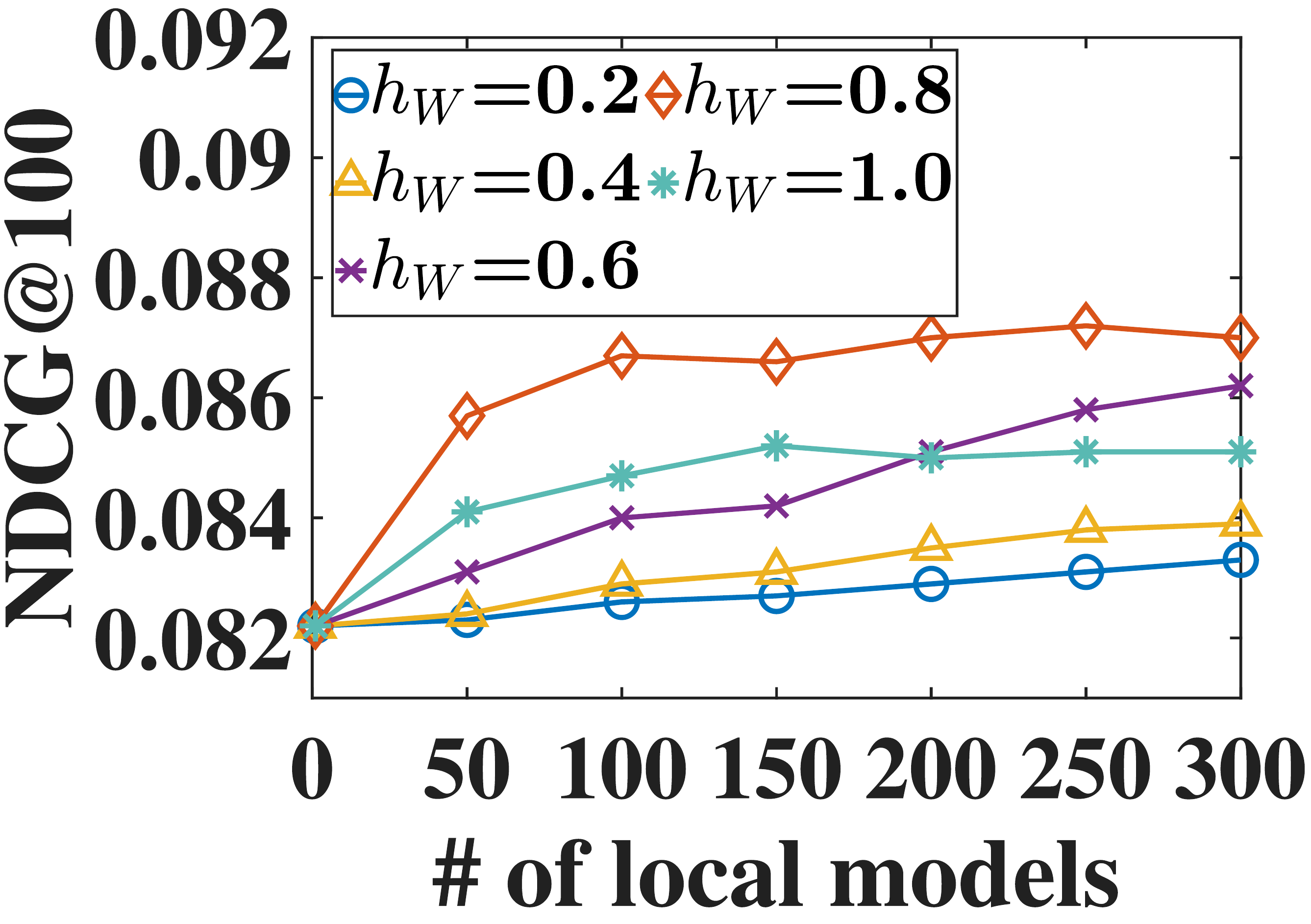} \\
(a) ML10M & (b) AMusic \\
\end{tabular}
\vskip -0.1in
\caption{NDCG@100 vs. kernel bandwidth $h_W$ in aggregating local models. ($h_T=1.0$)}
\label{fig:aggregate_h}
\vskip -0.1in
\end{figure}

%% file: Tab3_anchor.tex
\begin{table}[t] \small
\caption{NDCG@100 for different anchor selection methods with varying the number of local models. Coverage is our anchor selection method.}\label{tab:anchorselection}
\vspace{-0.1cm}
\begin{tabular}{c|c|ccc}
\toprule
\multirow{2}{*}{Dataset}    & Anchor    & \multicolumn{3}{c}{Number of local models} \\
                            & selection & 100 & 200 & 300 \\
\midrule
\multirow{4}{*}{ML10M}  & Random                 & 0.2030 (89.3\%) & 0.2052 (95.9\%) & 0.2068 (97.6\%)  \\
                         & K-means      & 0.2034 (92.9\%) & 0.2061 (97.1\%) & 0.2072 (98.6\%)  \\
                         & Farthest     & 0.2005 (83.1\%) & 0.2028 (93.4\%) & 0.2042 (96.9\%)  \\
                         & Coverage               & \textbf{0.2046 (97.3\%)} & \textbf{0.2062 (99.4\%)} & \textbf{0.2073 (99.9\%)} \\
\hline
\multirow{4}{*}{AMusic}  & Random                 & 0.0849 (53.5\%) & 0.0859 (64.8\%) & 0.0878 (71.3\%)  \\
                         & K-means       & 0.0856 (59.3\%) & 0.0871 (71.3\%) & 0.0882 (77.4\%)  \\
                         & Farthest               & 0.0835 (13.3\%) & 0.0853 (20.4\%) & 0.0865 (29.2\%)  \\
                         & Coverage               & \textbf{0.0865 (74.9\%)} & \textbf{0.0883 (83.6\%)} & \textbf{0.0893 (88.6\%)} \\
\bottomrule
\end{tabular}
\vspace{-0.2cm}
\end{table}

%% file: Tab4_case_study.tex
\efboxsetup{linecolor=red,linewidth=1pt}
\fboxsep=0mm

\begin{table}[t] \small
\caption{Illustration of top-5 recommendations for user 66005 in ML10M by LOCA$_{VAE}$. Out of the many local models, two were used to provide the final results. The red box indicates that the recommendation is in the ground truth.}
\label{tab:casestudy}
\centering
\begin{tabular}{>{\centering\arraybackslash}m{1.1cm} | >{\centering\arraybackslash}m{1cm} >{\centering\arraybackslash}m{1cm} >{\centering\arraybackslash}m{1cm} >{\centering\arraybackslash}m{1cm} >{\centering\arraybackslash}m{1cm} }
\toprule
Model               &  Top-1    & Top-2 & Top-3 & Top-4 & Top-5\\ 
\midrule
Local 70    & \setstretch{0.6} \includegraphics[width=1cm]{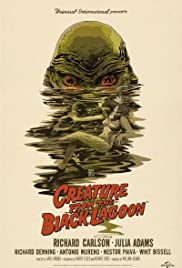}
              {\scriptsize \color{orange} Sci-Fi}{\scriptsize , Adventure}
            & \setstretch{0.6} \includegraphics[width=1cm]{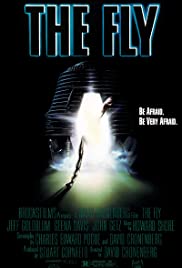}
              {\scriptsize \color{orange} Sci-Fi}{\scriptsize , \color{blue} Horror}
            & \setstretch{0.6} \efbox{\includegraphics[width=0.95cm]{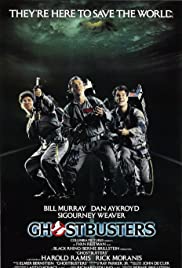}}
              {\scriptsize \color{orange} Sci-Fi}{\scriptsize , Action}
            & \setstretch{0.6} \includegraphics[width=1cm]{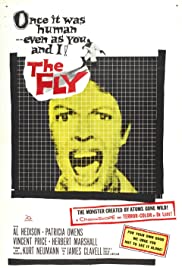}
              {\scriptsize \color{orange} Sci-Fi}{\scriptsize , \color{blue} Horror}
            & \setstretch{0.6} \includegraphics[width=1cm]{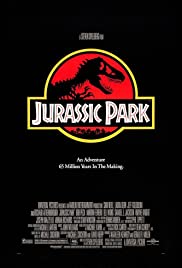} 
              {\scriptsize \color{orange} Sci-Fi}{\scriptsize , Action}  \\ \hline
Local 179   & \setstretch{0.6} \efbox{\includegraphics[width=0.95cm]{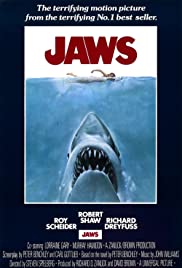}}
              {\scriptsize \color{blue} Horror}{\scriptsize , Action}
            & \setstretch{0.6} \efbox{\includegraphics[width=0.95cm]{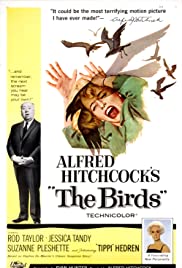}}
              {\scriptsize \color{blue} Horror}{\scriptsize , Drama}
            & \setstretch{0.6} \includegraphics[width=1cm]{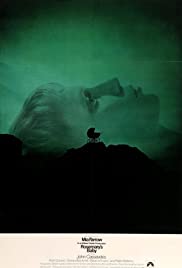}
              {\scriptsize \color{blue} Horror}{\scriptsize , Drama}
            & \setstretch{0.6} \includegraphics[width=1cm]{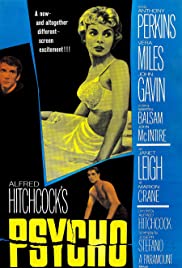}
              {\scriptsize \color{blue} Horror}{\scriptsize , Thriller}
            & \setstretch{0.6} \includegraphics[width=1cm]{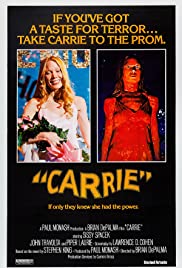}
              {\scriptsize \color{blue} Horror}{\scriptsize , Drama}\\ \hline
Global      & \setstretch{0.6} \includegraphics[width=1cm]{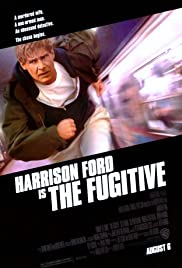}
              {\scriptsize Thriller, Action}
            & \setstretch{0.6} \includegraphics[width=1cm]{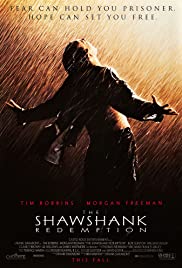}
              {\scriptsize Drama \vspace{1pt}}
            & \setstretch{0.6} \includegraphics[width=1cm]{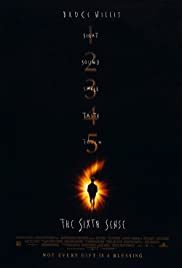}
              {\scriptsize Drama, Mystery}
            & \setstretch{0.6} \includegraphics[width=1cm]{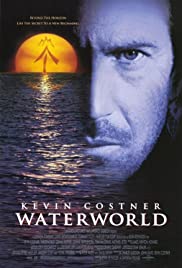}
              {\scriptsize \color{blue} Horror}{\scriptsize , Action}
            & \setstretch{0.6} \includegraphics[width=1cm]{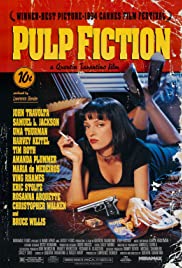}
              {\scriptsize Comedy, Crime} \\ \hline
Ground truth       & \setstretch{0.6} \includegraphics[width=1cm]{Poster/Ghostbustersa.k.a.GhostBusters.jpg}
              {\scriptsize \color{orange} Sci-Fi}{\scriptsize , Action}
            & \setstretch{0.6} \includegraphics[width=1cm]{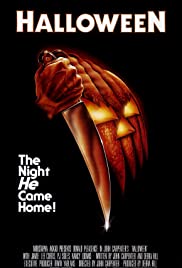}
              {\scriptsize \color{blue} Horror}{\scriptsize , Thriller}
            & \setstretch{0.6} \includegraphics[width=1cm]{Poster/Jaws.jpg}
              {\scriptsize \color{blue} Horror}{\scriptsize , Action}
            & \setstretch{0.6} \includegraphics[width=1cm]{Poster/BirdsThe.jpg}
              {\scriptsize \color{blue} Horror}{\scriptsize , Drama}
            & \setstretch{0.6} \includegraphics[width=1cm]{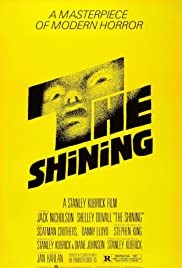} 
              {\scriptsize \color{blue} Horror}{\scriptsize , Thriller} \\
\bottomrule
\end{tabular}
\end{table}

%% file: sec-relatedwork.tex
\section{Related Work}\label{sec:relatedwork}

\subsubsection*{Local Latent Factor Models}
\label{sec:related_work_local}

The idea of using multiple local models was studied in ~\cite{BeutelCCPA17, XuBCC12, GeorgeM05}. In this study, we mainly discuss the local latent factor models under the local low-rank assumption. The existing local latent factor models can be categorized into two types. First, LLORMA~\cite{LeeKLS13} utilized multiple local matrices that represented local aspects of users, where local models overlapped with each other. The local collaborative ranking~\cite{LeeBKLS14} extended this idea to a pair-wise ranking objective. ~\cite{WangPJSW16} proposed a local latent factor model using WMF~\cite{HuKV08} under the same architecture. Although these models are capable of capturing local factors, they did not estimate global factors for all the users. Therefore, local models may not cover some users among the entire set of users. 

Second, GLSLIM~\cite{Christakopoulou16} proposed the estimation of a global model and multiple local item-item models, where the sub-matrices for local models were disjoint. sGLSVD~\cite{Christakopoulou18} used the same architecture with SVD~\cite{CremonesiKT10} as the base model. Compared to the existing local latent factor models, we assert that LOCA can be viewed as a generalized framework built upon the AE-based model.

\subsubsection*{Neural Recommender Models}

Deep neural networks (DNNs) have been widely used for recommender systems~\cite{ZhangYS17aa}. According to the model applicability, existing studies developed neural recommender models using various building blocks such as AE~\cite{SedhainMSX15,WuDZE16, WangWY15}, MLP~\cite{Cheng0HSCAACCIA16,CovingtonAS16,HeHSLJC18,Wang0WFC19}, CNN~\cite{HeDWTTC18,KimPOLY16}, RNN~\cite{HidasiKBT15,WuABSJ17}, and so on.

In this study, we mainly focus on AE-based models that are widely used to overcome the drawbacks of the existing linear recommender models, \ie, probabilistic matrix factorization (PMF)~\cite{SalakhutdinovM07}. As a  pioneering work, AutoRec~\cite{SedhainMSX15} adopted the conventional AE, that directly takes the user or item rating vectors as input and reconstructs the rating vector at the output layer. Unlike AutoRec, collaborative filtering neural network (CFN)~\cite{Strub15} was built upon stacked DAE to make it more robust. Collaborative denoising autoencoder (CDAE)~\cite{WuDZE16} adopted DAE by adding latent user vectors. MultVAE~\cite{LiangKHJ18} utilized the multinomial likelihood on top of VAE. Furthermore, EASE$^R$~\cite{Steck19} proposed the one-layer autoencoder and developed an analytic solution by relaxing the regularization of SLIM~\cite{NingK11}. Most recently, RecVAE~\cite{ShenbinATMN20} improved on MultVAE by using a composite prior distribution for the latent codes under the $\beta$-VAE framework. 

\input{Tab5_train_time}



%% file: Tab5_train_time.tex
\begin{table}
\small
\caption{Total training time (in seconds) of MultVAE~\cite{LiangKHJ18} and LOCA for varying the number of local models. }\label{tab:traintime}
\vskip -0.1in
\begin{tabular}{c|c|rrrrrr}
\toprule
\multirow{2}{*}{Dataset} & \multirow{2}{*}{MultVAE} & \multicolumn{6}{c}{Number of local models in LOCA$_{VAE}$}                                   \\
                 &       & $1$  & $2$   & $5$   & $10$ & $15$  & $20$ \\ 
\midrule
ML10M            & 949   & 248  & 308   & 410   & 726   & 1,101 & 1,459 \\ 
ML20M            & 2,514 & 427  & 450   & 482   & 1,167 & 1,495 & 1,710 \\
AMusic           & 46    & 18   & 27    & 32    & 53    & 76    & 82 \\
AGames           & 212   & 36   & 42    & 48    & 77    & 92    & 118 \\
Yelp             & 369   & 133  & 111   & 154   & 223   & 284   & 352 \\
\bottomrule 
\end{tabular}
\vskip -0.15in
\end{table} 


%% file: sec-conclusion.tex
\section{Conclusion}
\label{sec:conclusion}

This work presents a generalized local factor model, namely \emph{Local Collaborative Autoencoders (LOCA)}. To our knowledge, it is the first generalized framework under the local low-rank assumption that builds on the neural recommendation models. We explore a large number of local models by adopting a generalized framework with different weight schemes for training and aggregating them. Besides, we develop a novel method of discovering a sub-community to maximize the coverage of local models. Our experimental results demonstrate that LOCA is highly scalable, achieving state-of-the-art results by outperforming existing AE-based and local latent factor models on several large-scale public benchmarks.

\section*{Acknowledgment}

This work was supported by the National Research Foundation of Korea (NRF) (NRF-2018R1A5A1060031). Also, this work was supported by Institute of Information \& communications Technology Planning \& evaluation (IITP) grant funded by the Korea government (MSIT) (No.2019-0-00421, AI Graduate School Support Program and IITP-2020-0-01821, ICT Creative Consilience Program).

%% file: sec-appendix.tex
\newpage
\appendix